\documentclass{optica-article}

\journal{opticajournal}

\articletype{Research Article}

\usepackage{lineno}

\usepackage{booktabs}
\usepackage{siunitx}
\usepackage{longtable}
\usepackage{caption}
\usepackage{subcaption}
\usepackage{hyperref}
\usepackage{comment}
\usepackage[flushleft]{threeparttable}
\usepackage{tablefootnote}
\usepackage[normalem]{ulem}

\DeclareMathOperator{\Ci}{Ci}
\DeclareMathOperator{\Si}{Si}
\DeclareMathOperator{\Var}{Var}

\newcommand{\beginsupplement}{%
        \setcounter{table}{0}
        \renewcommand{\thetable}{S\arabic{table}}%
        \setcounter{figure}{0}
        \renewcommand{\thefigure}{S\arabic{figure}}%
     }

\begin{document}

\title{Coordinated international comparisons between optical clocks connected via fiber and satellite links}


\author{
Thomas~Lindvall,\authormark{1,\dag}
Marco~Pizzocaro,\authormark{2,\dag}
Rachel~M.~Godun,\authormark{3,*}
Michel~Abgrall,\authormark{4}
Daisuke~Akamatsu,\authormark{5, 6}
Anne~Amy-Klein,\authormark{7}
Erik~Benkler,\authormark{8}
Nishant~M.~Bhatt,\authormark{8}
Davide~Calonico,\authormark{2}
Etienne~Cantin,\authormark{7}
Elena~Cantoni,\authormark{2}
Giancarlo~Cerretto,\authormark{2}
Christian~Chardonnet,\authormark{7}
Miguel~Angel~Cifuentes Marin,\authormark{4}
Cecilia~Clivati,\authormark{2}
Stefano~Condio,\authormark{2}
E.~Anne~Curtis,\authormark{3}
Heiner~Denker,\authormark{9}
Simone~Donadello,\authormark{2}
Sören~Dörscher,\authormark{8}
Chen-Hao~Feng,\authormark{3}
Melina~Filzinger,\authormark{8}
Thomas~Fordell,\authormark{1}
Irene~Goti,\authormark{2}
Kalle~Hanhijärvi,\authormark{1}
H.~Nimrod~Hausser,\authormark{8}
Ian~R.~Hill,\authormark{3}
Kazumoto~Hosaka,\authormark{5}
Nils~Huntemann,\authormark{8}
Matthew~Y.H.~Johnson,\authormark{3}
Jonas~Keller,\authormark{8}
Joshua~Klose,\authormark{8}
Takumi~Kobayashi,\authormark{5}
Sebastian~Koke,\authormark{8}
Alexander~Kuhl,\authormark{8}
Rodolphe~Le~Targat,\authormark{4}
Thomas~Legero,\authormark{8}
Filippo~Levi,\authormark{2}
Burghard~Lipphardt,\authormark{8}
Christian~Lisdat,\authormark{8}
Hongli~Liu,\authormark{8}
Jérôme~Lodewyck,\authormark{4}
Olivier~Lopez,\authormark{7}
Maxime~Mazouth-Laurol,\authormark{4}
Tanja E.~Mehlstäubler,\authormark{8, 10}
Alberto~Mura,\authormark{2}
Akiko~Nishiyama,\authormark{5}
Tabea~Nordmann,\authormark{8}
Adam~O.~Parsons,\authormark{3}
Gérard~Petit,\authormark{11}
Benjamin~Pointard,\authormark{4}
Paul-Eric~Pottie,\authormark{4}
Matias~Risaro,\authormark{2}
Billy~I.~Robertson,\authormark{3}
Marco~Schioppo,\authormark{3}
Haosen~Shang,\authormark{4}
Kilian~Stahl,\authormark{8}
Martin~Steinel,\authormark{8}
Uwe~Sterr,\authormark{8}
Alexandra~Tofful,\authormark{3}
Mads~Tønnes,\authormark{4}
Dang-Bao-An~Tran,\authormark{3}
Jacob~Tunesi,\authormark{3}
Anders~E.~Wallin,\authormark{1}
and Helen~S.~Margolis\authormark{3}}

\address{

\authormark{1}VTT Technical Research Centre of Finland Ltd, National Metrology Institute VTT MIKES, P.O.~Box 1000, FI-02044 VTT, Finland\\

\authormark{2}Istituto Nazionale di Ricerca Metrologica (INRIM), Strada delle Cacce 91, 10135, Torino, Italy\\

\authormark{3}National Physical Laboratory (NPL), Hampton Road, Teddington, TW11 0LW, U.K.\\

\authormark{4}Laboratoire Temps Espace (LNE-OP), Observatoire de Paris, Université PSL, Sorbonne Université, Université de Lille, LNE, CNRS, 61 avenue de l'Observatoire, 75014 Paris, France\\

\authormark{5}National Metrology Institute of Japan (NMIJ), National Institute of Advanced Industrial Science and Technology (AIST), 1-1-1 Umezono, Tsukuba, Ibaraki 305-8563, Japan\\

\authormark{6}Department of Physics, Graduate School of Engineering Science, Yokohama National University, 79-5 Tokiwadai, Hodogaya-ku, Yokohama 240-8501, Japan\\

\authormark{7}Laboratoire de Physique des Lasers (LPL), Université Sorbonne Paris Nord, CNRS, 99 Avenue Jean-Baptiste Clément, 93430 Villetaneuse, France\\

\authormark{8}Physikalisch-Technische Bundesanstalt (PTB), Bundesallee 100, 38116 Braunschweig, Germany\\

\authormark{9}Institut für Erdmessung, Leibniz Universität Hannover (LUH), Schneiderberg 50, 30167 Hannover, Germany\\

\authormark{10}Leibniz Universität Hannover (LUH), Welfengarten 1, 30167 Hannover, Germany\\

\authormark{11}Bureau International des Poids et Mesures (BIPM), 92312 Sèvres Cedex, France

\authormark{\dag} joint first authors\\
}
\email{\authormark{*}rachel.godun@npl.co.uk} 



\begin{abstract*}

Optical clocks provide ultra-precise frequency references that are vital for international metrology as well as for tests of fundamental physics.  To investigate the level of agreement between different clocks, we simultaneously measured the frequency ratios between ten optical clocks in six different countries, using fiber and satellite links.  This is the largest coordinated comparison to date, from which we present a subset of 38 optical frequency ratios and an evaluation of the correlations between them.  Four ratios were measured directly for the first time, while others had significantly lower uncertainties than previously achieved, supporting the advance towards a redefinition of the second and the use of optical standards for international time scales.

\end{abstract*}

\section{Introduction}

Frequency is the physical quantity that can be measured more precisely than any other. This places optical atomic clocks, with uncertainties at the 17th and 18th digits, among the best tools for probing fundamental physics such as general relativity \cite{Delva2017, Sanner2019}, variations of fundamental constants \cite{Lange2021, Sherrill2023} and searches for dark matter \cite{Roberts2020, Beloy2021, Kobayashi2022, Filzinger2023}, as well as applications such as relativistic geodesy \cite{Chou2010, Grotti2018, Takamoto2020}. Optical clocks are also candidates for redefining the second in the International System of Units (SI) \cite{Dimarcq2024}, replacing the lower-accuracy Cs-based standards that are currently used, and enabling the transition to optical time scales \cite{Grebing2016, Hachisu2018, Yao2019, Formichella2024}.

Today, a growing number of optical clocks are available worldwide, based on several different atomic species. If operated together, they could provide an extremely powerful asset for the applications above. However, clocks in different countries are mostly compared via satellite techniques, which introduce uncertainties up to 100 times larger than from the clocks themselves, thus compromising the effort. Frequency distribution with optical fibers has already enabled non-local comparisons with the links contributing lower uncertainties than the clocks ($ {<} 1 \times 10^{-18}$). So far, however, only a few accurate comparisons with more than two optical clocks running simultaneously have been carried out, at best involving clocks in 2 or 3 different locations and sometimes revealing discrepancies greater than the estimated uncertainties \cite{Beloy2021, Takamoto2015, Riedel2020, Amy-Klein2024}.

Measurement campaigns comparing at least 3 clocks simultaneously and employing more than one link technique can be far more insightful than the mostly pairwise comparisons carried out to date.  Although significantly more complex to carry out, coordinated campaigns with multiple clocks enable a large number of optical frequency ratios to be measured simultaneously.  This allows consistency checks, enabling systems that are not operating correctly to be identified and eliminated, which is not possible with only two clocks and a single link.

The need for more extended clock comparison campaigns in view of the redefinition of the second, targeted for 2030, has also been recognized by the international metrology community, which has defined a roadmap with mandatory criteria that must be achieved~\cite{Dimarcq2024}. The criterion that is currently the least well advanced is the validation of optical frequency standard uncertainty budgets, which requires more clock comparisons to demonstrate that systems are operating within their expected uncertainties.

Here, we report on the largest coordinated international comparison of optical clocks to date.  As shown in Fig.~\ref{fig:Uptimes}(a), ten optical clocks in six different countries were compared simultaneously.   The frequency comparisons were carried out over optical fiber and satellite links, and an overview of the  measurement campaign is presented in section~\ref{sec:overview}.
 The data analysis is described in sections~\ref{sec:fibre_comparisons} and \ref{sec:satellite_comparisons}, with the results discussed in more detail in section~\ref{sec:results_discussion}.  When presenting frequency ratios derived from multiple clocks running simultaneously, it is also important to take account of correlations because the ratios are not entirely independent of each other.  Section~\ref{sec:correlations} considers the sources of correlations in the results and evaluates the correlation coefficients between different frequency ratios.  The main conclusions are summarized in section~\ref{sec:conclusion}.

In addition to providing datasets for tests of fundamental physics, we anticipate that the results from such a large-scale comparison will be a much-needed addition to the body of international clock comparison data. This is used in the calculation of recommended values for standard frequencies \cite{Margolis2024}, allowing optical frequency standards to contribute to International Atomic Time (TAI) as secondary representations of the SI second.  Furthermore, the uncertainties and consistency of measured frequency ratios will influence the choice of which optical transition(s) should be used in the new definition of the SI second.

\section{Overview of the coordinated international comparison}
\label{sec:overview}

\begin{figure}[p]
    \begin{subfigure}{\textwidth}
    \caption{}
    \centering
    \includegraphics[width=0.5\textwidth]{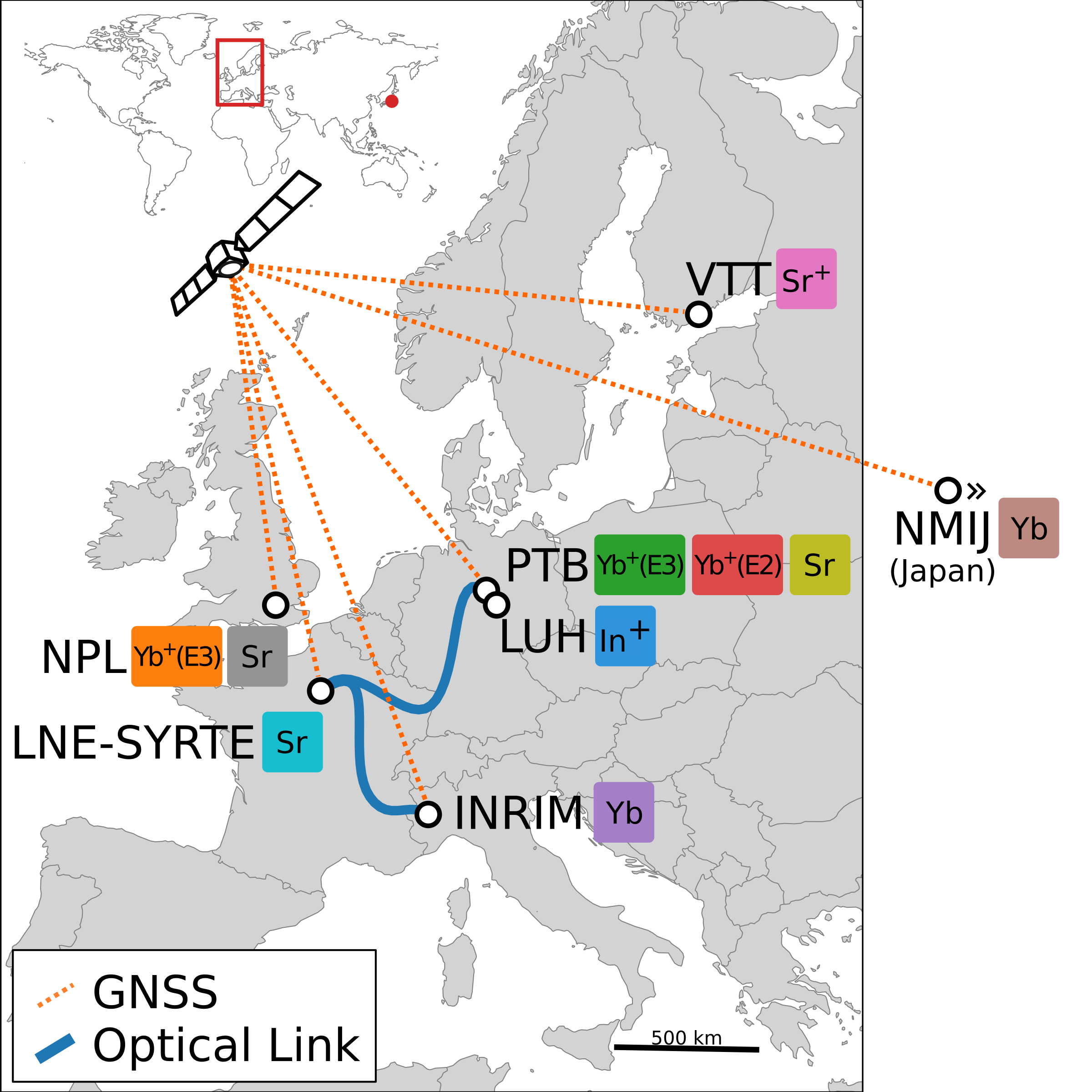}
    \end{subfigure}
    \begin{subfigure}{\textwidth}
    \caption{}
    \centering
    \includegraphics[width=0.7\textwidth]{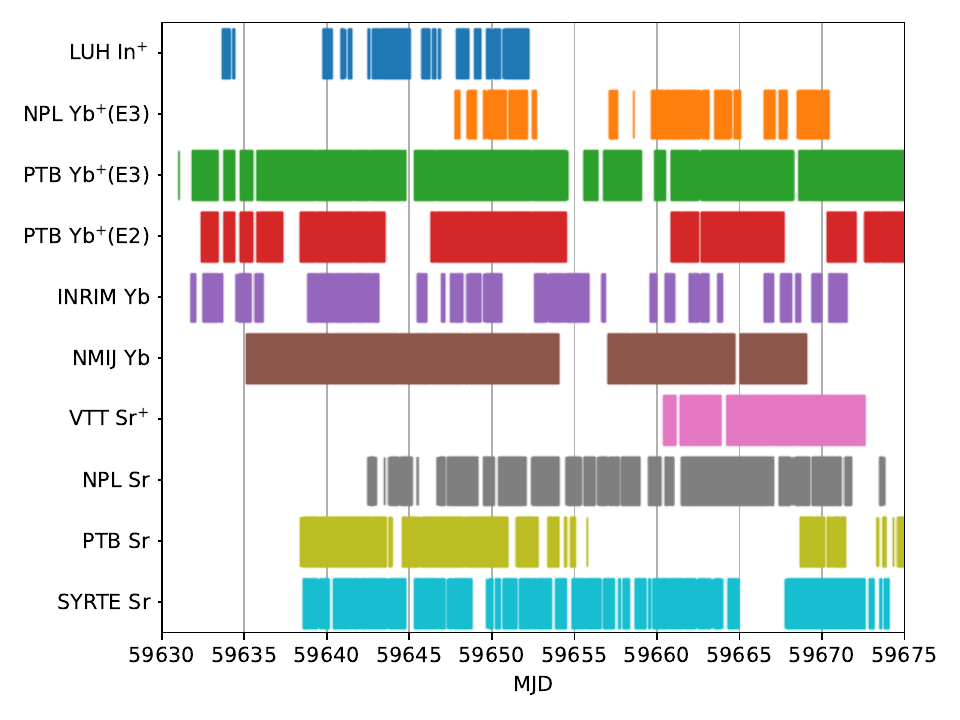}
    \end{subfigure}
    \begin{subfigure}{\textwidth}
    \caption{}
    \centering
    \includegraphics[width=0.7\textwidth]{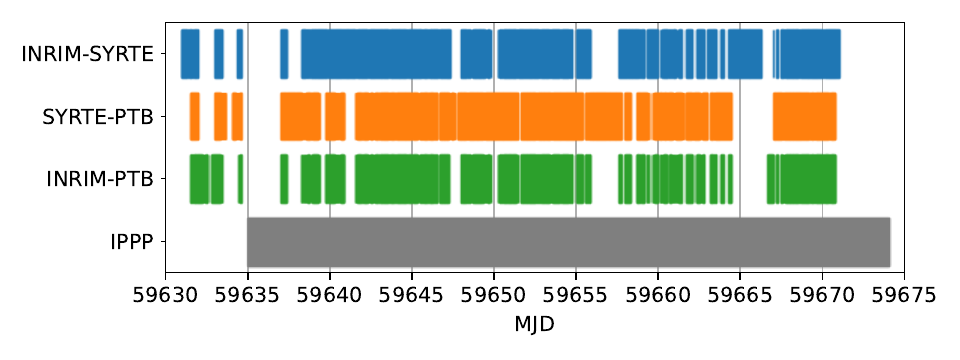}
    \end{subfigure}

    \caption{Overview of the clock comparison campaign: ten optical clocks in six different countries were compared over 45 days. (a) Map of the links and geographical distribution of the clocks. Optical fiber links (blue solid lines) connect LNE-SYRTE with INRIM (1023 km) and LNE-SYRTE with PTB (1370 km), resulting also in a connection between INRIM and PTB. All institutes are connected by GNSS (orange dotted lines). The LUH In$^+$ clock is located on the PTB campus. Local comparisons are carried out between the 2 clocks at NPL and between the 4 clocks at PTB/LUH. (b) Uptimes of the clocks during the campaign. (c) Uptime of the  international fiber links and the IPPP evaluation period for all the comparisons made via GNSS links. Uptimes are represented as colored regions as a function of time from Modified Julian Date (MJD) 59630 (20th of February 2022) to MJD 59675 (6th of April 2022).}
    \label{fig:Uptimes}
\end{figure}

As part of a European collaborative project, ROCIT~\cite{Margolis_2024b}, a coordinated comparison of optical clocks was carried out over 45 days in 2022 involving partners in Finland, France, Germany, Italy and the UK, alongside collaborators in Japan, see Fig.~\ref{fig:Uptimes}(a).  Table~\ref{tab:clocks} shows the ten optical clocks that were compared, along with the estimated systematic uncertainties for each clock.  The lowest clock uncertainties, $u_\mathrm{B}$, were at the fractional frequency level of just a few $10^{-18}$.
Ideally,  frequency transfer links that do not introduce any appreciable uncertainty into the measurements should be chosen and so, where possible, the clocks were compared either locally (at NPL and PTB) or else connected via international optical fibers that have been shown to support comparisons at the $10^{-18}$ level and below~\cite{Lisdat2016, Koke2019, Cantin2021, Clivati2022}.  For comparisons between clocks where fiber links were not available, the frequency transfer was carried out via Integer Precise Point Positioning (IPPP) \cite{Petit2015}, making use of Global Positioning System (GPS) data. However, since the IPPP technique is applicable to any Global Navigation Satellite System, the general term GNSS is used in the following. The operational times with valid data, also known as uptimes, for all the clocks and links are presented in Fig.~\ref{fig:Uptimes}(b) and (c), showing 45 days starting from the Modified Julian Day (MJD) 59630 (20th February 2022).

\begin{table}[tb]
\centering
\begin{threeparttable}
\caption{Clocks participating in the measurement campaign. For each clock, the means of comparison is shown as well as the estimated fractional uncertainties ($10^{-18}$) associated with systematic frequency shifts of the clock, $u_\mathrm{B}$, the relativistic redshift correction to the reference potential $W_0$, $u_\mathrm{RRS}$ \cite{Denker2018,Riedel2020}, and the radio frequency (rf) distribution chain for GNSS comparison, $u_\mathrm{rf}$ (not relevant for fiber or local comparisons).
\label{tab:clocks}}
\footnotesize
\begin{tabular}{llllccccc}
\toprule
Institute           &Clock                  &Identifier     &Link               & $u_\mathrm{B}$ & $u_\mathrm{RRS}$ & $u_\mathrm{rf}$ & Ref. \\
\midrule
INRIM, Italy        & $^{171}$Yb            & IT-Yb1        & Fiber \& GNSS     & 20    & 2.7   & 30    & \cite{Goti2023}\\
LNE-SYRTE,$^\dag$ France   & $^{87}$Sr      & SYRTE-Sr2     & Fiber \& GNSS     & 17    & 3.0   & 58    & \cite{Lodewyck2016}\\
LUH,$^\ddag$ Germany& $^{115}$In$^+$        & PTB-In1              & Fiber \& Local             & 2.5   & 2.4   & {-}   & \cite{Hausser2025}\\
NMIJ, Japan         & $^{171}$Yb            & NMIJ-Yb1      & GNSS              & 110   & 6.0   & 100   & \cite{Kobayashi2020, Kobayashi2022}\\
NPL, UK             & $^{87}$Sr             & NPL-Sr1       & GNSS \& Local              & 22    & 2.7   & 49    & \cite{Hobson2020}\\
NPL, UK             & $^{171}$Yb$^+$(E3)    & NPL-E3Yb+3    & GNSS \& Local             & 3.2   & 2.5   & 49    & \cite{Tofful2024}\\
PTB, Germany        & $^{87}$Sr             & PTB-Sr3       & Fiber, Local \& GNSS     & 3.0   & 2.4   & 10    &  \cite{Schwarz2022}\\
PTB, Germany        & $^{171}$Yb$^+$(E2)    & PTB-Yb1E2     & Local             & 26    & {-}   & {-}   & \cite{Lange2021} \\
PTB, Germany        & $^{171}$Yb$^+$(E3)    & PTB-Yb1E3     & Fiber, Local \& GNSS     & 2.7   & 2.4   & 10    & \cite{Sanner2019}\\
VTT, Finland        & $^{88}$Sr$^+$         & MIKES-Sr+1    & GNSS              & 10    & 2.4   & 10    & \cite{Lindvall2022} \\
\bottomrule
\end{tabular}
\begin{tablenotes}
\item{ $^\dag$ LNE-SYRTE is now called LTE (Laboratoire Temps-Espace/LNE-OP).}
\item{ $^\ddag$ LUH clock is located on the PTB campus.}
\end{tablenotes}
\end{threeparttable}
\end{table}

\section{Clock comparisons via fiber link}
\label{sec:fibre_comparisons}

The European network of phase-stabilized fiber links connects the optical clocks at NPL in the UK, SYRTE in France, PTB in Germany and INRIM in Italy \cite{Lisdat2016, Cantin2021, Clivati2022, Schioppo2022}.
The network comprises thousands of kilometers of optical fibers, partly shared with internet traffic, in a star-like topology with the Paris area acting as a central node. For this comparison, the SYRTE-PTB and SYRTE-INRIM links were operational, with connections at the French-German and French-Italian borders in Strasbourg and Modane. The network enabled clocks at PTB and INRIM to be compared for the first time via a fiber link, which is 2370 km long. In France, the network uses the national research infrastructure REFIMEVE~\cite{Cantin2021,REFIMEVE} and, in Italy, it uses the Italian Quantum Backbone (IQB) national network~\cite{Clivati2020a}. The SYRTE-NPL link was not available at this time due to missing connections in both the London and Paris sections of the fiber. Each link in the network is equipped with repeater laser stations\cite{Guillou-Camargo2018} and bidirectional erbium or Brillouin amplifiers for signal recovery\cite{Koke2019}. Each span of the link is characterized by establishing a second independent link.  The uncertainties introduced by the optical links reach ${<}\num{1e-18}$ after 1000\;s of averaging time and are negligible relative to that of the clocks.

Data from the optical clocks and fiber links were recorded every 1 s, synchronized to a local realization of Coordinated Universal Time (UTC), with the uptimes shown in Figure \ref{fig:Uptimes}. The average frequency ratios were calculated as the mean over all valid data points and the results are shown in Table~\ref{Tab:FrequencyRatios}. The data recording and analysis followed the universal formalism introduced by Lodewyck \emph{et al.}~\cite{Lodewyck2020}.

The uncertainties on the frequency ratios in Table~\ref{Tab:FrequencyRatios} include both the statistical and systematic uncertainties, with the systematic contributions from each clock shown in Table~\ref{tab:clocks}. Since the clocks have different heights in the Earth's gravity potential, it is necessary to take account of the relativistic redshift (RRS) of the clock frequencies~\cite{Petit2005}. For remote clock comparisons, the differential shift is obtained from differencing the RRS of the two clocks, which are measured relative to an absolute reference potential that is close to sea level and defined to be $W_0=62~636~856.00$~m$^2$s$^{-2}$~\cite{Riedel2020,Denker2018}. The uncertainty of the differential shift is obtained from the values $u_\mathrm{RRS}$ for each clock, as shown in Table 1. For local clock comparisons, the differential shift is derived from the height difference, which can be measured more directly and its uncertainty is smaller. For PTB Yb$^+$(E2), which is involved only in the ratio against PTB Yb$^+$(E3), the $u_\mathrm{RRS}$ uncertainty is omitted because there is no relativistic redshift as both transitions were measured in the same ion.

The statistical uncertainties were obtained from the observed white-frequency noise contribution to the Allan deviation that was  evaluated for each ratio. We observed day-to-day scatter in some frequency ratios greater than expected based on the white noise contribution alone. To account for this, we calculated the Birge ratio (square root of the reduced chi-squared) from the averages in daily bins (see the Supplementary Material for details). As is commonly done, the statistical uncertainty was then inflated by the Birge ratio.

\begin{table}[htbp]
\caption{Summary of the frequency ratios measured in this campaign, shown with the estimated uncertainties for each measurement.  It is likely, however, that some of the frequency ratios have significantly larger uncertainties than the estimates shown here.  See Section~\ref{sec:results_discussion} for further discussion of discrepancies seen in the ratios measured via GNSS with INRIM as well as ratios involving SYRTE Sr and PTB Sr.
}
\label{Tab:FrequencyRatios}
\footnotesize\rm
\begin{tabular}{@{}*{15}{l}}
\hline
No. &Frequency ratio with total       &Total fractional       &Link  &Clock 1    &Clock 2\\
    &uncertainty in parentheses       &uncertainty            &      &           &\\
\hline
1   &1.973~773~591~557~215~789(9)  &$4.4 \times 10^{-18}$  &Local  &LUH In$^+$    &PTB Yb$^+$(E3)\\
2   &2.445~326~324~126~950~199(58)  &$2.4 \times 10^{-17}$  &Fiber  &LUH In$^+$    &INRIM Yb\\
3   &2.952~748~749~874~860~909(15)  &$5.1 \times 10^{-18}$  &Local  &LUH In$^+$    &PTB Sr\\
4   &2.952~748~749~874~861~332(72)  &$2.4 \times 10^{-17}$  &Fiber  &LUH In$^+$    &SYRTE Sr\\
5   &1.072~007~373~634~205~468(29)  &$2.7 \times 10^{-17}$  &Local  &PTB Yb$^+$(E2)&PTB Yb$^+$(E3)\\
6   &1.238~909~231~832~259~569(26)  &$2.1 \times 10^{-17}$  &Fiber  &PTB Yb$^+$(E3)&INRIM Yb\\
7   &1.495~991~618~544~900~525(36)  &$2.4 \times 10^{-17}$  &Local  &NPL Yb$^+$(E3)&NPL Sr\\
8   &1.495~991~618~544~900~659(8)   &$5.4 \times 10^{-18}$  &Local  &PTB Yb$^+$(E3)&PTB Sr\\
9   &1.495~991~618~544~900~897(32)  &$2.1 \times 10^{-17}$  &Fiber  &PTB Yb$^+$(E3)&SYRTE Sr\\
10  &1.207~507~039~343~337~793(26)  &$2.2 \times 10^{-17}$  &Fiber  &INRIM Yb   &PTB Sr\\
11  &1.207~507~039~343~337~981(36)  &$2.9 \times 10^{-17}$  &Fiber  &INRIM Yb   &SYRTE Sr\\
12  &1.000~000~000~000~000~146(21)  &$2.1 \times 10^{-17}$  &Fiber  &PTB Sr     &SYRTE Sr\\
13  &0.999~999~999~999~999~80(28)   &$2.8 \times 10^{-16}$  &GNSS   &NPL Yb$^+$(E3)&PTB Yb$^+$(E3)\\
14  &1.238~909~231~832~259~82(37)   &$3.0 \times 10^{-16}$  &GNSS   &NPL Yb$^+$(E3)&INRIM Yb\\
15  &1.238~909~231~832~259~18(45)   &$3.6 \times 10^{-16}$  &GNSS   &NPL Yb$^+$(E3)&NMIJ Yb\\
16  &1.238~909~231~832~260~04(11)   &$8.8 \times 10^{-17}$  &GNSS   &PTB Yb$^+$(E3)&INRIM Yb\\
17  &1.238~909~231~832~259~60(20)   &$1.6 \times 10^{-16}$  &GNSS   &PTB Yb$^+$(E3)&NMIJ Yb\\
18  &1.443~686~489~498~354~68(51)   &$3.5 \times 10^{-16}$  &GNSS   &NPL Yb$^+$(E3)&VTT Sr$^+$\\
19  &1.443~686~489~498~354~89(17)   &$1.2 \times 10^{-16}$  &GNSS   &PTB Yb$^+$(E3)&VTT Sr$^+$\\
20  &1.495~991~618~544~900~59(56)   &$3.7 \times 10^{-16}$  &GNSS   &NPL Yb$^+$(E3)&PTB Sr\\
21  &1.495~991~618~544~900~66(48)   &$3.2 \times 10^{-16}$  &GNSS   &NPL Yb$^+$(E3)&SYRTE Sr\\
22  &1.495~991~618~544~900~51(25)   &$1.7 \times 10^{-16}$  &GNSS   &PTB Yb$^+$(E3)&NPL Sr\\
23  &1.495~991~618~544~900~94(15)   &$1.0 \times 10^{-16}$  &GNSS   &PTB Yb$^+$(E3)&SYRTE Sr\\
24  &0.999~999~999~999~999~65(18)   &$1.8 \times 10^{-16}$  &GNSS   &INRIM Yb   &NMIJ Yb\\
25  &1.165~288~345~913~157~59(18)   &$1.6 \times 10^{-16}$  &GNSS   &INRIM Yb   &VTT Sr$^+$\\
26  &1.165~288~345~913~158~03(31)   &$2.7 \times 10^{-16}$  &GNSS   &NMIJ Yb    &VTT Sr$^+$\\
27  &1.207~507~039~343~337~30(23)   &$1.9 \times 10^{-16}$  &GNSS   &INRIM Yb   &NPL Sr\\
28  &1.207~507~039~343~337~33(13)   &$1.1 \times 10^{-16}$  &GNSS   &INRIM Yb   &PTB Sr\\
29  &1.207~507~039~343~337~52(16)   &$1.3 \times 10^{-16}$  &GNSS   &INRIM Yb   &SYRTE Sr\\
30  &1.207~507~039~343~337~74(33)   &$2.7 \times 10^{-16}$  &GNSS   &NMIJ Yb    &NPL Sr\\
31  &1.207~507~039~343~337~82(21)   &$1.8 \times 10^{-16}$  &GNSS   &NMIJ Yb    &PTB Sr\\
32  &1.207~507~039~343~338~03(24)   &$2.0 \times 10^{-16}$  &GNSS   &NMIJ Yb    &SYRTE Sr\\
33  &1.036~230~254~578~831~95(24)   &$2.4 \times 10^{-16}$  &GNSS   &VTT Sr$^+$    &NPL Sr\\
34  &1.036~230~254~578~832~29(26)   &$2.5 \times 10^{-16}$  &GNSS   &VTT Sr$^+$    &PTB Sr\\
35  &1.036~230~254~578~832~33(21)   &$2.0 \times 10^{-16}$  &GNSS   &VTT Sr$^+$    &SYRTE Sr\\
36  &1.000~000~000~000~000~08(24)   &$2.4 \times 10^{-16}$  &GNSS   &NPL Sr     &PTB Sr\\
37  &1.000~000~000~000~000~10(23)   &$2.3 \times 10^{-16}$  &GNSS   &NPL Sr     &SYRTE Sr\\
38  &1.000~000~000~000~000~14(12)   &$1.2 \times 10^{-16}$  &GNSS   &PTB Sr     &SYRTE Sr\\
\hline
\end{tabular}
\end{table}

\section{Clock comparisons via satellite link}
\label{sec:satellite_comparisons}

A key challenge in comparing clocks via IPPP links is that continuous phase measurements are needed in order to average down the link noise as $1/T$, where $T$ is the measurement time. The optical clock data, however, contained gaps of varying lengths as seen from Fig.~\ref{fig:Uptimes}. Hydrogen masers (HMs) were therefore used as flywheels and were compared continuously over the IPPP link during the selected analysis intervals.
The optical-clock vs maser frequency ratios were then evaluated and extrapolated across optical clock downtimes, which introduces an additional uncertainty contribution.
This extrapolation uncertainty was evaluated using the Fourier transform method \cite{Dawkins2007}, where the uncertainty is obtained from the modelled power spectral density of the maser and the Fourier transform of a weighting function that depends on the uptime of the clock and the analysis interval. For this, a noise model for each HM used for extrapolation was estimated using optical clock vs HM data from the campaign and/or prior information, see Supplementary Material.

The exact measurement configuration varied between institutes. At VTT, the optical clock was measured against a free-running HM, which was also used as a reference for the GNSS receiver. In the other institutes, the receivers were referenced to the local UTC($k$) realization, and if the optical clock was measured against a free-running HM, an additional HM-UTC($k$) measurement was used to complete the frequency chain between the clock and the receiver. All data were provided in the 30\;s binned format of the GNSS RINEX (Receiver Independent Exchange Format) files and IPPP solutions.

When a clock was measured against a free-running HM, the frequency at the center of the analysis interval was obtained by correcting the mean frequency using the HM drift and the difference between the interval center and the barycenter of the data. The drift was estimated from a linear fit to all valid clock vs HM data.

The Bureau International des Poids et Mesures (BIPM) carries out IPPP processing for several of the involved receivers on a regular basis, but additional receivers were used in the analysis for this measurement campaign to allow a common-clock receiver comparison for each institute. This enabled issues with phase steps or excursions in the solutions to be revealed. By comparing both receivers at a given institute against a remote receiver/HM, one can also identify which receiver suffered the phase issue. Where available, a receiver without issues was selected for the analysis. Otherwise, care was taken not to extrapolate over phase issues. Similarly, extrapolation over steep phase ramps in the masers was avoided. If varying the analysis interval changed the frequency ratio by a significant fraction (${\gtrsim}50\%$) of the estimated extrapolation uncertainty, this was taken as a sign of undesirable maser behaviour and extrapolation was avoided.

For IPPP, a link frequency transfer uncertainty (FTU) of $1\times 10^{-15}/(T/\mathrm{d})$ was used, where $T$ is the time measured in days (d). This is an empirical, conservative estimate that has been validated for up to ${\sim}100$ days \cite{Leute2018,Petit2021,Petit2022}. It includes the effect of typical temperature coefficients of the GNSS equipment as well as unidentifiable systematic effects. If a piece of equipment has a particularly large temperature coefficient, this would be seen as diurnals in the IPPP solution, which was not observed here. Under optimal circumstances, i.e., with good receivers and smooth tropospheric variations, the link FTU can be slightly lower, but this is not known \emph{a priori}. On the other hand, worse performance can usually be identified from the quality of the IPPP solution. For this campaign, a higher FTU of $1.3\times 10^{-15}/(T/\mathrm{d})$ was used for links involving the NM0D receiver (used by NMIJ during the first half of the campaign) due to its larger residual daily boundary phase steps. This value was estimated from a common-clock comparison between the two NMIJ receivers.

The total statistical uncertainty was evaluated by adding in quadrature the contributions from the IPPP link, the maser extrapolation, the statistical uncertainties of the clocks, and the maser drift uncertainties. The first two typically dominated the total uncertainty, and the analysis intervals and thus the amount of maser extrapolation were varied separately for each ratio to minimize the total statistical uncertainty. As many of the clocks had at least one longer gap in the data and, in some cases, there were receiver and maser anomalies to avoid, most of the frequency ratios were evaluated as a weighted mean of two analysis intervals. For simplicity, HM and IPPP link correlations between the two intervals for a particular ratio, separated by at least one day and in all but one case by more than two days, were neglected to allow a regular weighted mean to be used. The analysis of correlations, see Sec.~\ref{sec:correlations}, justified this approach.
In all cases where two intervals were used, the two ratios agreed within their combined statistical uncertainty.
Due to the higher statistical uncertainty of the IPPP ratios, no additional statistical methods such as Birge ratios were needed for these comparisons.  Systematic uncertainties, as shown in Table~\ref{tab:clocks}, were also included in the total uncertainties. This includes the uncertainty from the radio frequency (rf) distribution chain, $u_\mathrm{rf}$.

\section{Discussion of measured frequency ratios}
\label{sec:results_discussion}
\begin{figure}[t]
    \centering
    \includegraphics[width=1\textwidth]{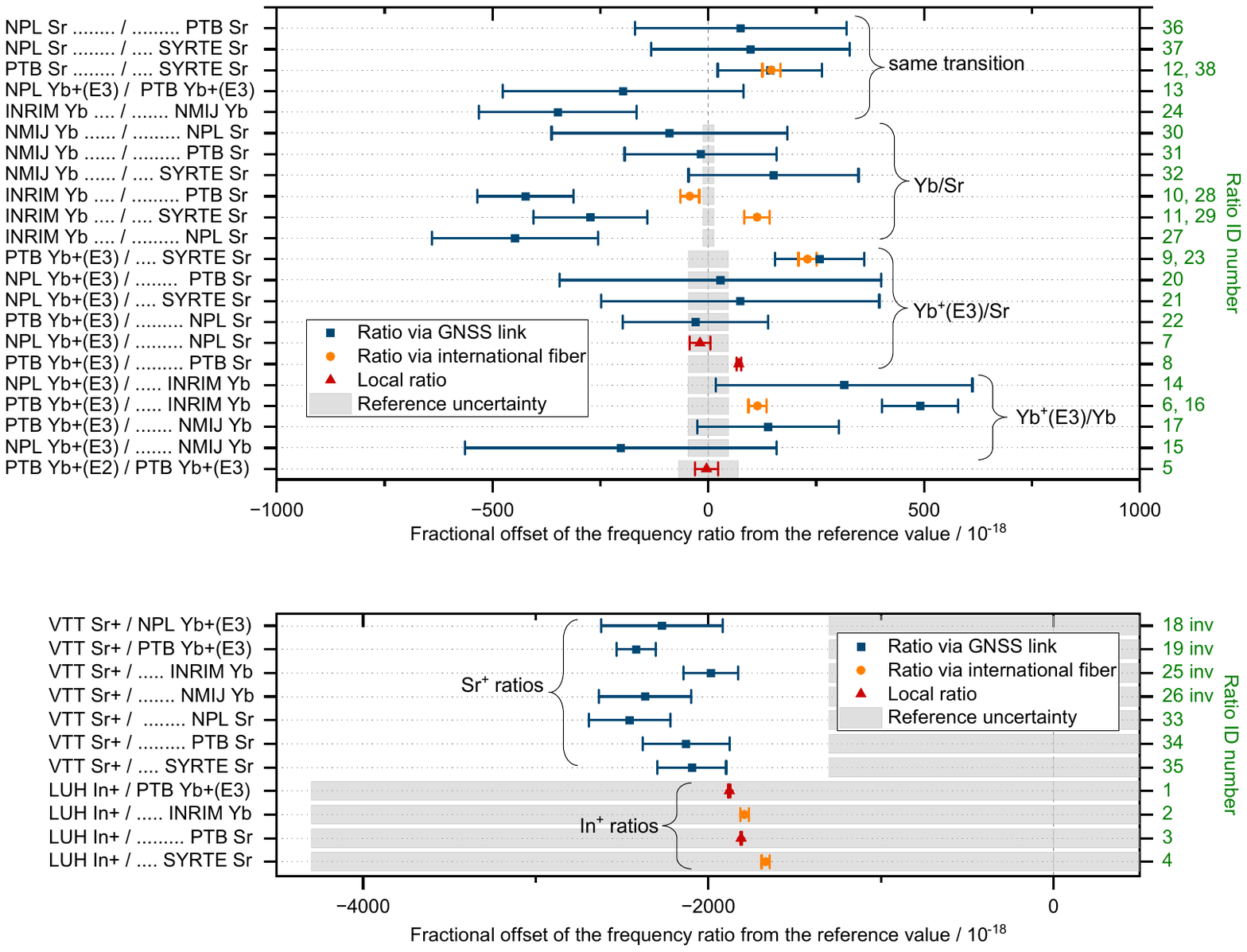}
    \caption{Frequency ratios measured in March 2022 via GNSS links (blue squares), international fiber links (orange circles) and local comparisons (red triangles). The error bars on the data points represent the total relative standard uncertainties for each measurement, including statistical and systematic uncertainties.  The grey bars show the relative standard uncertainties on the reference frequency ratios, which are obtained from the least-squares adjustment of standard frequencies, approved by the International Committee for Weights and Measures (CIPM) in 2021 and reported in Table~S1 of the Supplementary Material.  The right-hand axis of the plots shows the ratio ID numbers, corresponding to the rows in Table~\ref{Tab:FrequencyRatios} that give the measured values of the frequency ratios; the ID numbers are marked with `inv' if the ratio is inverted between the graph and the table.}
    \label{fig:results_overview}
\end{figure}

The frequency ratios measured during this campaign are listed in Table~\ref{Tab:FrequencyRatios} and also shown in Fig.~\ref{fig:results_overview}.
Each measured frequency ratio is plotted in the graph as the fractional offset from a reference value.
The reference values are taken from ~\cite{Margolis2023} and are the result of a least-squares adjustment carried out in 2021~\cite{Margolis2024}, similar to the process used for the recommended values of physical constants \cite{Tiesinga2021,Margolis2015}. The input of the adjustment consisted of all absolute frequency and frequency ratio measurements published at the time.  A full list of the optimized values that are used as the reference frequency ratios in this paper is given, along with the corresponding uncertainties, in the Supplementary Material.
We note that several of the reference ratios are dominated by a single measurement with lower uncertainty than the other measurements contributing to the reference value of that ratio. For ratios that have not been directly measured, the reference ratio can then be dominated by two such single measurements.

For this clock comparison campaign, we have chosen not to present all the measured frequency ratios between every pair of clocks in the network.  Instead, we present a subset of 38 ratios, in groups that demonstrate consistencies and identify outliers.  Each of these groups will be discussed in more detail in the sections that follow.  The remaining frequency ratios and their associated uncertainties can all be derived from the measurements presented here, taking into account the correlations between them.

\subsection{Fiber vs GNSS ratios}
Five frequency ratios were measured via both fiber and GNSS links. The ratios between SYRTE Sr and the two PTB clocks show good agreement between the two link technologies. However, GNSS-derived frequency ratios involving the INRIM Yb clock show a discrepancy of $4\times 10^{-16}$ compared to optical link results, likely caused by an unidentified problem in the signal distribution at INRIM, also observed in a comparison with Cs fountains \cite{Clivati2022}.  Repeated tests conducted over the following months consistently confirmed nominal performance of the INRIM equipment and the origin of the problem in March 2022 has not been identified. Further comparisons between the optical link and the GNSS link will be useful to identify the issue or confirm that it is resolved. This serves to illustrate the importance of carrying out large, coordinated measurement campaigns with multiple clocks and links running simultaneously in order to identify and eliminate such inconsistencies.  For this particular measurement campaign, we therefore consider the results of all the frequency ratios via GNSS to INRIM to be unreliable.

\subsection{Same-transition comparisons}
The expected frequency ratio for same-transition comparisons is 1 with no uncertainty, allowing for a clear check of whether the clocks and links are behaving as expected. This campaign allowed for same-transition comparisons between pairs of clocks based on Sr, Yb and Yb$^+$(E3), as shown in Fig.~\ref{fig:results_overview}.

Three different Sr/Sr frequency ratios were measured via GNSS links and the results show agreement between the Sr clocks at NPL, PTB and SYRTE within 1--2 standard uncertainties.  As is the case for several of the ratios measured via GNSS in this campaign, the uncertainty on the comparison between PTB Sr and SYRTE Sr is below \num{1.8e-16}, thus improving upon the best previously achieved uncertainty in a satellite comparison~\cite{Riedel2020}.
However, the even smaller uncertainty on the fiber link comparison reveals a fractional frequency difference of \num{1.46(21)e-16} between the PTB Sr and SYRTE Sr clocks.
Looking at all the ratios involving the SYRTE Sr clock shows similar offsets and a larger than expected level of scatter in the SYRTE clock's frequency during this campaign (see Supplementary Material for details), indicating an uncontrolled frequency shift at the \num{e-16} level. Moreover, as will be discussed in the following subsections, the PTB Sr clock may also have had issues during this campaign.
Discrepancies between Sr clocks have also been seen in other measurement campaigns~\cite{Amy-Klein2024, Grotti2024} but the coordinated set of frequency ratios recorded here provides further insight.

The GNSS measurements show the NPL and PTB Yb$^+$(E3) clocks agreeing within the combined relative standard uncertainty of $2.7 \times 10^{-16}$, dominated by the maser extrapolation at NPL during downtime of the optical clock.

The Yb/Yb frequency ratio was measured directly between the clocks at INRIM and NMIJ via GNSS.  As the measurement involved the GNSS equipment at INRIM, the result is considered unreliable.  One can, however, evaluate the ratio between the Yb clocks at INRIM and NMIJ by combining the frequency ratios of each clock measured relative to a third clock, connected to INRIM by fiber link.  Choosing PTB Yb$^+$(E3) as the third clock because of its high uptime, $(\text{INRIM Yb} / \text{PTB Yb}^+\text{(E3)}) \times (\text{PTB Yb}^+\text{(E3)} / \text{NMIJ Yb}) - 1 = 2(17) \times 10^{-17}$, demonstrating agreement between the two Yb clocks at the level of $1.7 \times 10^{-16}$.

\subsection{Different-transition comparisons}
Frequency ratios between different clock transitions are not known \emph{a priori} so we
use the reference frequency ratios, derived in the least squares adjustment carried out in 2021 \cite{Margolis2024}, as the expected values.

\subsubsection{Yb/Sr}
There were two Yb clocks (at INRIM and NMIJ) and three Sr clocks (at NPL, PTB and SYRTE) in this campaign.  Measurements involving NMIJ Yb were carried out via GNSS link and the three Yb/Sr frequency ratios involving NMIJ Yb all agree with the reference values within one relative standard uncertainty ($\sim 2 \times 10^{-16}$).
Ignoring the measurements involving the GNSS equipment at INRIM leaves two further direct measurements of the Yb/Sr frequency ratio by fiber link. Both are offset from the reference frequency ratio by more than one standard deviation of their combined relative uncertainties, with the difference between the two ratios consistent with the discrepancy between the Sr clocks at PTB and SYRTE, as mentioned above.

\subsubsection{Yb$^+$(E3)/Sr}
Both NPL and PTB operated Yb$^+$(E3) and Sr clocks in this measurement campaign.
Considering also the Sr clock running at SYRTE, seven values of the Yb$^+$(E3)/Sr frequency ratio were obtained.
The PTB Yb$^+$(E3) and SYRTE Sr clocks were linked via both GNSS and fiber and the results show good agreement between the two different link techniques but both ratios are more than two relative standard uncertainties above the reference frequency ratio.  This is again consistent with the SYRTE Sr clock frequency being too low during this measurement campaign. The three other Yb$^+$(E3)/Sr ratios derived via GNSS links are in good agreement with the reference frequency ratio.  There are also two local measurements of Yb$^+$(E3)/Sr, one at NPL and the other at PTB, which are of particular interest since the measurements did not involve long-distance links between the clocks, nor any uncertainty associated with the gravity potential difference between NPL and PTB.
The NPL measurement is consistent with the reference value, which is based largely on earlier results from PTB~\cite{Doerscher2021} using a different Sr clock from the one used here.
The PTB Yb$^+$(E3)/Sr frequency ratio measured in this campaign is not consistent with either of these results at the \num{e-17} level.
It was concluded in \cite{Hausser2025}, using data from further measurements, that the PTB Sr clock frequency was likely a few $10^{-17}$ too low during this campaign.
However, this does not explain the discrepancies in the Sr/Sr and Yb/Sr frequency ratios discussed above, as those would increase even further if a correction were applied.
It is impossible to conclude unambiguously from the measurements in this campaign alone whether the offset is with the PTB Sr clock or with the reference values, so we refrain from applying additional frequency corrections or uncertainties (unlike Ref. \cite{Hausser2025}).
Repeated measurement campaigns are needed in order to gather more data to contribute to the least-squares optimization process and reduce the uncertainty in the reference values.

\subsubsection{Yb$^+$(E3)/Yb}
This campaign marks the first direct measurements of the Yb$^+$(E3)/Yb frequency ratio.
If we ignore measurements involving the GNSS connection to INRIM, we have three measures of the Yb$^+$(E3)/Yb frequency ratio---two via GNSS link to NMIJ Yb and one via fiber link to INRIM Yb. The two results via GNSS link are consistent with the reference frequency ratio, whereas the lower uncertainty result via fiber link is offset from the reference value by approximately twice the combined relative uncertainty of \num{5e-17}. Given that the reference value has approximately twice the uncertainty of the value measured via fiber link, this measurement will be able to have a significant influence over the future optimized value for this ratio.

\subsubsection{Yb$^+$(E2)/Yb$^+$(E3)}
The frequency ratio Yb$^+$(E2)/Yb$^+$(E3) measured locally at PTB agrees with the reference value, well within the combined uncertainties. The PTB Yb$^+$(E2) clock was operated at the same time as the PTB Yb$^+$(E3) clock and the two share the same physics package. Given that the uptimes for PTB Yb$^+$(E2) are highly overlapped with PTB Yb$^+$(E3), ratios of the PTB Yb$^+$(E2) clock with other clocks in the network can be obtained by combination of ratios with  PTB Yb$^+$(E3).

\subsection{Ratios involving Sr$^+$ and In$^+$}
For both Sr$^+$ and In$^+$, the 2021 recommended frequency values have uncertainties above $10^{-15}$ and all ratios involving these species have correspondingly large uncertainties on their reference values.  The ratios involving Sr$^+$ or In$^+$ in this campaign have therefore been plotted on separate axes in Fig.~\ref{fig:results_overview} with a larger scale.

It can be seen that all the measured frequency ratios involving Sr$^+$ are consistently offset from the reference values by a little over $2 \times 10^{-15}$.  However, the Sr$^+$ ratios measured here are in agreement with recent results in the published literature~\cite{Steinel2023, Jian2023}.  This strongly suggests that the recommended frequency value for the Sr$^+$ secondary representation of the second is offset from the unperturbed transition frequency by approximately twice its assigned uncertainty of $\num{1.3e-15}$. At the time of the 2021 least-squares adjustment, no optical frequency ratios involving Sr$^+$ had been published, and the recommended frequency value is strongly dominated by a single absolute frequency measurement \cite{Barwood2014}, which in the light of recent results is to be considered suspect. To date, only a single optical frequency ratio (Sr$^+$/Yb$^+$(E3)) has been published \cite{Steinel2023}, so this campaign has produced the first direct measurements of the Sr$^+$/Sr and Sr$^+$/Yb ratios and
will thus be able to contribute to an improved optimized value in the future.

The frequency ratios involving In$^+$ in this measurement campaign are all consistent with their respective reference values, but with much lower uncertainties.  We therefore expect that the data presented here, along with related measurements in \cite{Hausser2025}, will allow the In$^+$ recommended frequency to be determined with much lower uncertainty in the next update to the recommended values of standard frequencies.  We also note that this campaign has made the first direct measurement of the In$^+$/Yb frequency ratio, with an uncertainty just over \num{2e-17}.

\section{Correlations between different frequency ratios}
\label{sec:correlations}

The frequency ratios measured during this campaign  depend on common input quantities (shared clock and link data) and are not all independent of each other.
Multivariate measurement models require, beyond the standard uncertainties, estimates of the covariance matrix \cite{BIPM_GUM, BIPM_GUM_S2}.
In our case, 38 frequency ratios require the calculation of a $38\times38$ covariance matrix or of 703 correlation coefficients.
Efforts were therefore made to identify the non-zero correlations and to recognize the largest common effects when measuring optical frequency ratios \cite{ROCIT_D3}.

The correlations between the results of this campaign are visualized in Fig. \ref{fig:FibreCorr} and reported in the Supplementary Material. We calculated correlations from the statistical and systematic uncertainties of each clock, including the RRS uncertainty, which was considered strongly correlated between clocks in the same location. For the GNSS ratios we also considered that measurements from the same institute share the RF distribution, the temporal correlation of the IPPP solutions, and the correlations from the extrapolation uncertainties calculated for the same physical HM.

For local and fiber link measurements, the largest contributions resulted from the systematic uncertainties of the clocks, which we considered fully correlated between measurements involving the same clock.  The correlations arising from the statistical noise of each clock depended on the temporal overlap between the two measurements \cite{ROCIT_D3}. They were calculated assuming white frequency noise and left unchanged by the Birge ratio expansion. The largest resulting correlation coefficients, up to 0.94 in magnitude, are between pairs of ratios carried out via fiber measurements with either INRIM Yb or SYRTE Sr as the common clock in the two ratios.

For the GNSS ratios, the extrapolation uncertainty is the source of the largest correlations. To evaluate these, we generalized the Fourier transform method used to calculate extrapolation uncertainties \cite{Dawkins2007} to calculate covariances (see Supplementary Material). These correlation coefficients have values up to 0.80 in magnitude.

Correlations from the systematic uncertainty of the clock, RF distribution, and RRS were significant in particular for ratios involving NMIJ, with correlation coefficients up to 0.75 in magnitude.
Correlations between ratios involving different clocks sharing the RF distribution and RRS uncertainty concerned only NPL and PTB and were relatively small, with correlation coefficients up to 0.05 in magnitude.

The IPPP link noise is dominated by flicker phase noise \cite{Petit2022} and thus introduces non-trivial correlation for measurements sharing GNSS receivers.
A model for the autocorrelation function of the IPPP link phase noise was calculated using data from an IPPP-fiber comparison and adjusted to agree with the frequency transfer uncertainty of $1\times 10^{-15}/(T/\mathrm{d})$ for intervals above 1~day (see Supplementary Material for details). The numerical values of the correlation coefficients from the IPPP links are up to 0.17 in magnitude.

Correlations between the frequency ratios obtained via the GNSS links and via the fiber links were calculated to be negligible in magnitude (${<} 0.01$) because the uncertainty introduced by the GNSS link is much larger than the statistical and systematic uncertainties of the clocks compared using both techniques.

Overall, the calculation of correlation coefficients allows us to consider the results collectively rather than in isolation. This will facilitate combining our results with future measurements, for checking the consistency of optical clocks or for the next calculation of recommended frequency values \cite{Margolis2024}, ensuring that they are unbiased and with properly estimated uncertainties.

\begin{figure}

    \begin{subfigure}{\textwidth}
    \caption{}
    \centering
    \includegraphics[width=0.4384\textwidth]{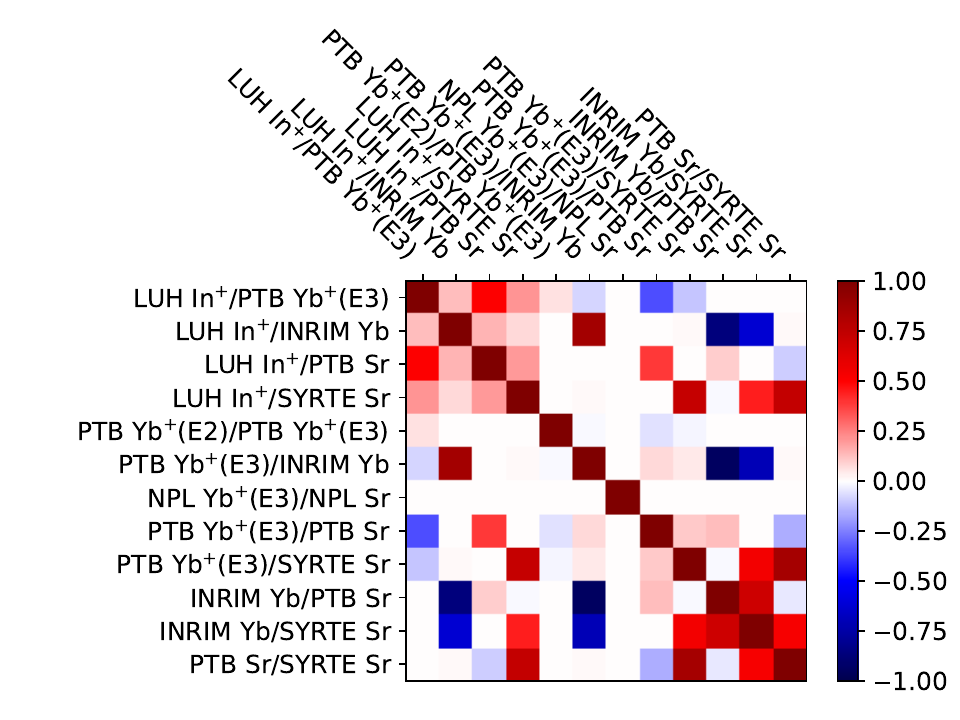}
    \end{subfigure}
    \begin{subfigure}{\textwidth}
    \caption{}
    \centering
    \includegraphics[width=0.75\textwidth]{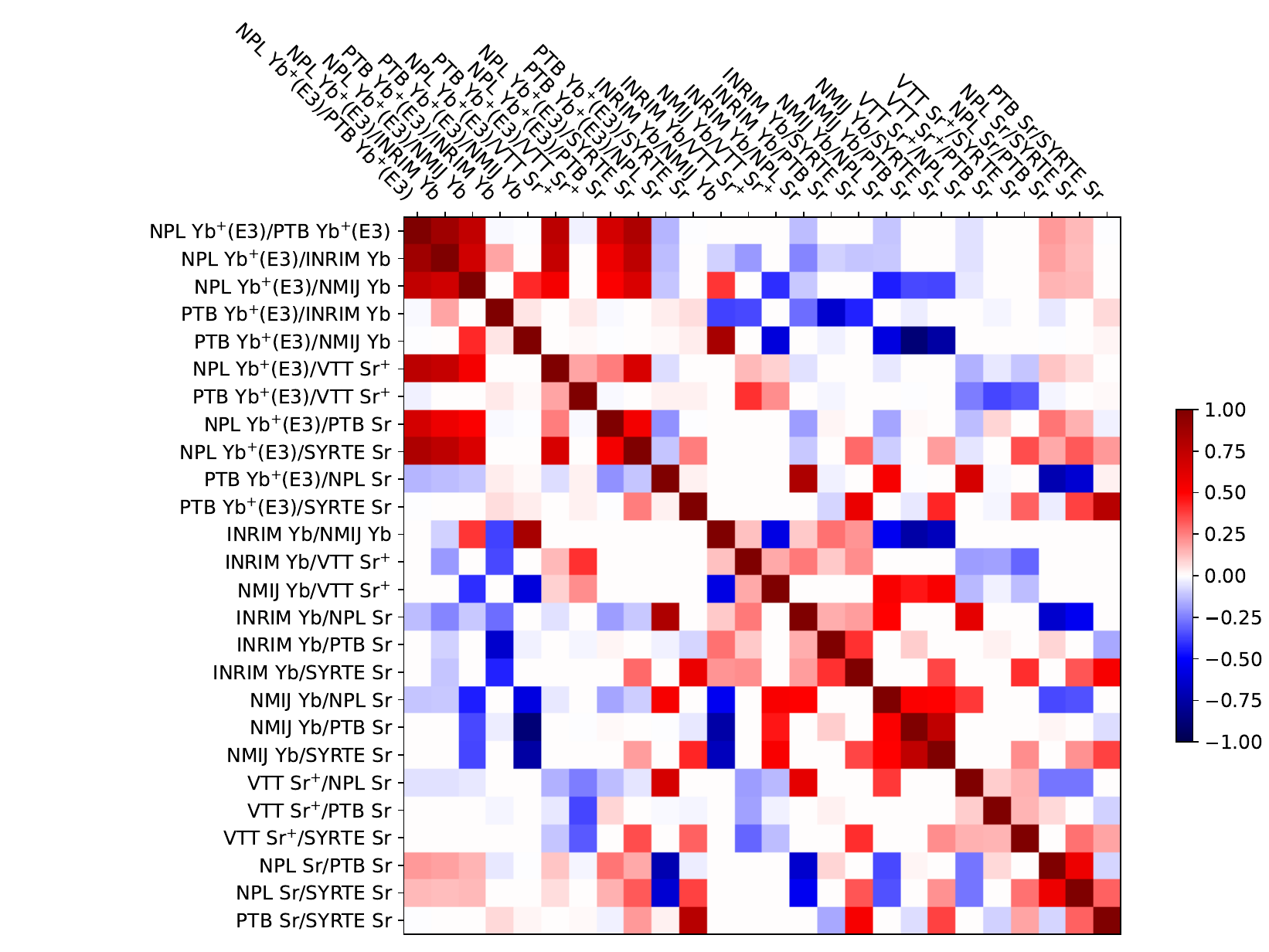}
    \end{subfigure}

    \caption{Graphical representation of the correlation values between the ratios reported in Table~\ref{Tab:FrequencyRatios}.  (a) Correlations in frequency ratios measured locally or via international fiber links.
    (b) Correlations in the frequency ratios measured via GNSS frequency transfer techniques.}
    \label{fig:FibreCorr}
\end{figure}

\section{Conclusions}
\label{sec:conclusion}
We have carried out the largest coordinated comparison of optical clocks to date, simultaneously comparing ten optical clocks in six different countries connected via fiber or satellite links.  We have presented 38 frequency ratios, including the first direct measurements of four optical frequency ratios: Yb$^+$(E3)/Yb, In$^+$/Yb, Sr$^+$/Sr and Sr$^+$/Yb.
Due to the length of the campaign, the high uptime of the clocks, and the use of IPPP, several of the GNSS ratios had total uncertainties below $1.8\times 10^{-16}$, the lowest uncertainty previously achieved in a satellite comparison.  Additionally, the frequency ratios involving the Sr$^+$ and In$^+$ clocks all had significantly lower uncertainties than the corresponding reference values based on the 2021 least-squares adjustment.
We also evaluated the correlation coefficients between all the measured frequency ratios, which required new analysis methods to deal with measurements that shared a common maser, and also to deal with measurements that relied on IPPP solutions from common GNSS data.  In total, 242 non-zero correlation coefficients were computed, with 155 of these having an absolute value greater than 0.1.

As we advance towards a redefinition of the SI second and international time scales based on optical standards, it becomes increasingly important to demonstrate the feasibility of operating a network of optical clocks.  We have demonstrated agreement between GNSS and optical fiber links over a continental scale by comparing frequency ratios measured by more than one link technique.  Furthermore, we have been able to verify many of the estimated measurement uncertainties by comparing frequency ratios that were measured by more than one pair of institutes. In some cases, the frequency ratios measured in this campaign have revealed inconsistencies, indicating where caution is required in the use of those results.  Specifically, we identified that during the period of these measurements, the signal distribution at INRIM introduced an offset of $\num{4e-16}$ into frequency ratios via GNSS.  As was previously reported, the Sr clock at PTB may also have had an offset at the level of a few $10^{-17}$. Comparisons between the frequency ratios in this campaign and the reference values also indicated a possible offset in the Sr clock at SYRTE of up to $\num{2e-16}$.  However, in some cases it is not clear from this campaign alone whether discrepancies relative to the reference values indicate an issue with the frequency ratios measured here, or an issue with the reference values that were themselves derived from other results in the literature.
More data from repeated measurement campaigns will be required to resolve such ambiguities.

We anticipate that the frequency ratios reported here will contribute to the global dataset of optical frequency ratios, allowing the accuracy of reference values to be improved in the future. Having better reference values will help to confirm when measurement systems are operating correctly and increase the confidence in the use of optical clocks for advancing metrology as well as fundamental physics.


\begin{backmatter}

\bmsection{Acknowledgment}
This work was carried out in the EMPIR project ROCIT 18SIB05.  This project has received funding from the EMPIR programme co-financed by the Participating States and from the European Union's Horizon 2020 research and innovation programme.
NPL acknowledges support by the UK government Department for Science Innovation and Technology through the National Measurement System Programme.
NMIJ acknowledges support by the Japan Society for the Promotion of Science (JSPS) KAKENHI Grant Number 17H01151, 17K14367, and 22H01241, and JST-Mirai Program Grant Number JPMJMI18A1, Japan.
PTB acknowledges support by the Deutsche Forschungsgemeinschaft (DFG, German Research Foundation) under Germany’s Excellence Strategy EXC-2123 QuantumFrontiers (Project-ID 390837967) and SFB 1227 DQ-mat (Project-ID 274200144, within project B02 and B03); this work was partially supported by the Max Planck-RIKEN-PTB Center for Time, Constants and Fundamental Symmetries funded equally by the three partners.
VTT	acknowledges support by the Academy of Finland (decisions 339821 and 328786).
The work at VTT is also part of the Academy of Finland Flagship Programme `Photonics Research and Innovation' (PREIN, decision 320168).
Optical fiber links from the French REFIMEVE network are supported by Program ``Investissements d’Avenir'' launched by the French Government and implemented by Agence Nationale de la Recherche with references ANR-11-EQPX-0039 (Equipex REFIMEVE+) and ANR-21-ESRE-0029 (ESR/Equipex+ T-REFIMEVE).
M. M.-L. received funding from the European Union's Horizon 2020 research and innovation programme under Grant Agreement No. 951886 (CLONETS-DS).
M. T. received funding from DIM SIRTEQ network of Conseil R\'egional \^Ile de France (ATH-2019 ONSEPA).
Authors from NMIJ are indebted to the colleagues of Time Standards Group and Optical Frequency Measurement Group at NMIJ, and also of Yokohama National University. NMIJ acknowledges the Geospatial Information Authority of Japan for the determination of the geopotential value of the Yb optical clock at NMIJ.

\bmsection{Disclosures}
The authors declare no conflicts of interest.

\bmsection{Data Availability Statement}
Data underlying the results presented in this paper are not publicly available at this time but may be obtained from the authors upon reasonable request.

\end{backmatter}



\newpage
\title{Supplementary Material}
\beginsupplement
\setcounter{section}{0}

\section{Reference frequency ratios based on CIPM 2021 recommended values}
\label{App:ref_vals}
The reference frequency ratios, as used in Fig.~2 in the main text, are given in Table~\ref{Tab:RefFreqRatios}.
The ratios are ordered in the table by highest frequency in the numerator clock, and then highest frequency in the denominator clock.
 The ratios were derived as part of a least-squares adjustment process to provide recommended values for standard frequencies, based on measurements in the published literature.  The recommended frequencies were last updated and approved by the International Committee for Weights and Measures (CIPM) in 2021 \cite{S-Margolis2024}.  Alongside the 2021 recommended frequencies, the least-squares adjustment process also provided an associated set of frequency ratios~\cite{S-Margolis2023}. It is these ratios, consistent with the 2021 recommended frequency values, that are used as the reference frequency ratios in this paper.

\begin{table}[h]
\centering
\caption{Reference frequency ratios, consistent with the 2021 recommended frequency values.  The values were obtained from the data repository~\cite{S-Margolis2023}, associated with~\cite{S-Margolis2024} and are used with the full precision shown here, while the fractional uncertainties are from~\cite{S-Margolis2024}.}
\label{Tab:RefFreqRatios}
\small\rm
\begin{tabular}{@{}*{15}{l}}
\hline
Atomic species                 &Reference frequency ratio                 &Fractional uncertainty\\
\hline
$^{115}$In$^+$/$^{171}$Yb$^+$(E3)   &1.973~773~591~557~219~495~298~33   &$4.3\times 10^{-15}$\\
$^{115}$In$^+$/$^{171}$Yb           &2.445~326~324~126~954~577~515~80   &$4.3\times 10^{-15}$\\
$^{115}$In$^+$/$^{87}$Sr                &2.952~748~749~874~866~252~596~06   &$4.3\times 10^{-15}$\\
$^{171}$Yb$^+$(E2)/$^{171}$Yb$^+$(E3)   &1.072~007~373~634~205~472~639~55   &$6.9\times 10^{-17}$\\
$^{171}$Yb$^+$(E3)/$^{171}$Yb           &1.238~909~231~832~259~427~891~89   &$4.7\times 10^{-17}$\\
$^{171}$Yb$^+$(E3)/$^{88}$Sr$^+$        &1.443~686~489~498~351~403~342~78   &$1.3\times 10^{-15}$\\
$^{171}$Yb$^+$(E3)/$^{87}$Sr            &1.495~991~618~544~900~552~345~03   &$4.6\times 10^{-17}$\\
$^{171}$Yb/$^{88}$Sr$^+$                &1.165~288~345~913~155~272~344~57   &$1.3\times 10^{-15}$\\
$^{171}$Yb/$^{87}$Sr                    &1.207~507~039~343~337~845~067~45   &$1.3\times 10^{-17}$\\
$^{88}$Sr$^+$/$^{87}$Sr                 &1.036~230~254~578~834~498~456~96   &$1.3\times 10^{-15}$\\
\hline
\end{tabular}
\end{table}

\section{Analysis of clock comparisons via fiber link}

The network was divided into several nodes (corresponding to optical clocks, optical cavities or fiber-disseminated lasers) and data was collected as comparisons between nodes in a compatible format \cite{S-ROCITformat} and shared on a common repository. The format handles data equivalently from optical clocks, optical cavities, optical combs and fiber links. Data were recorded on a 1-s grid. Each node was synchronized to a local realization of Coordinated Universal Time (UTC) to $\ll 1$~s. Missing and invalid data points were also flagged by each operator at this stage.  Data were combined to derive the optical ratios on the 1-s grid only on the common uptime of the chain.

The average ratios were calculated as the mean over all valid points on the 1-s grid.  To assess the statistical uncertainties, the frequency ratios were also averaged in daily bins, as shown in Fig.~\ref{fig:fibre_link_statistics(a)}(a).  This allowed the calculation of Birge ratios for each dataset, as indicated on the plots. The corresponding Allan deviations for each dataset are shown in Fig.~\ref{fig:fibre_link_statistics(b)}(b).

We note that Fig.~\ref{fig:fibre_link_statistics(a)}(a) shows for many ratios an excess daily scatter when uncertainties are calculated from the white noise contribution alone. This effect is also evident in the departure from white noise at long averaging times in the Allan deviations in Fig.~\ref{fig:fibre_link_statistics(b)}(b). To account for these effects we inflated the statistical uncertainty  by the Birge ratio. The choice of daily bins, despite each clock achieving or striving for continuous operation, is motivated by the typical timescale of human intervention on the various experiments. With the exception of ratios involving the SYRTE Sr clock, the observed Birge ratios range from 1 to 2.1 and are not dissimilar to values reported for other long-running clock comparison campaigns \cite{S-Beloy2021, S-Doerscher2021, S-Ohmae2020,  S-Ushijima2015, S-Baillard2008}. Birge ratios for comparisons with the SYRTE Sr clock were between 3.3 and 5.3, indicating a larger scatter as discussed in section 5 of the main paper.

We note that  the source of excess daily scatter may vary between the clocks involved.
In general, long-term deviations from white-frequency noise in the instabilities may be tied to systematic frequency shifts. For example, the observed instabilities are expected for measurements carried out in different conditions, systematic shifts that vary between days, or in the case of systematic corrections which are applied from daily measurements. We note that this last effect results in a scatter around the mean of the systematic correction even if the underlying systematic shift is constant.

\begin{figure}[h]
    \centering
    \begin{subfigure}{\textwidth}
    \caption{}
        \includegraphics[width=\textwidth]{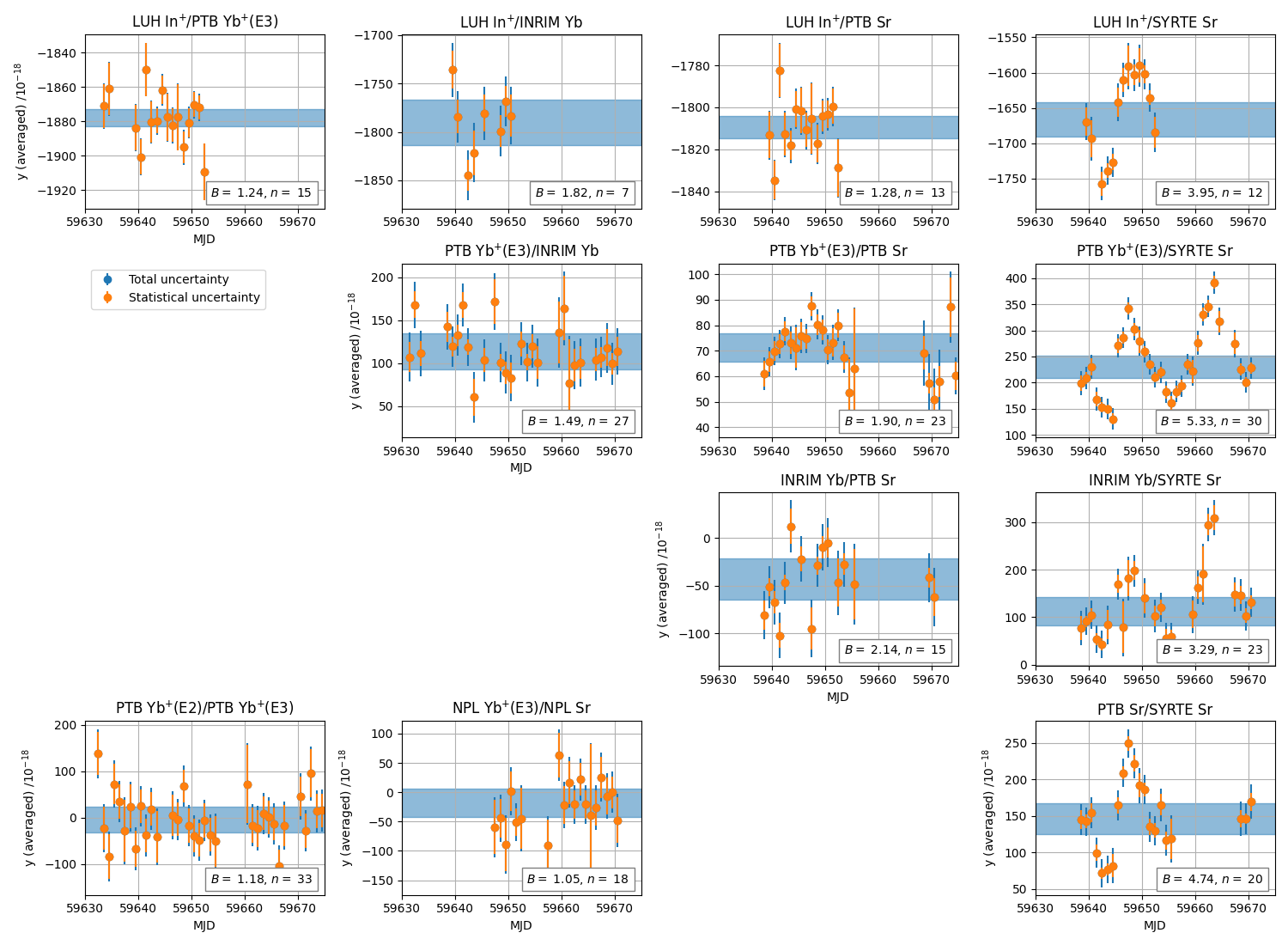}
    \end{subfigure}
    \caption{\textbf{(a)} Overview of the data from clock comparisons via fiber links or local measurements. Measured daily average ratios, presented as fractional differences from the reference ratios. The box in each plot reports the Birge ratio $B$ and the number of degrees of freedom $n$. The blue regions show the average value with the total uncertainty, evaluated as the sum in quadrature of the contributions from systematic effects (shown in Table 1 of the main text) and statistics (derived via the Allan deviations and Birge ratios as shown in Fig.~\ref{fig:fibre_link_statistics(b)}(b)).}
    \label{fig:fibre_link_statistics(a)}
\end{figure}
\clearpage

\begin{figure}[h]
    \ContinuedFloat
    \centering
    \begin{subfigure}{\textwidth}
    \caption{}
        \includegraphics[width=\textwidth]{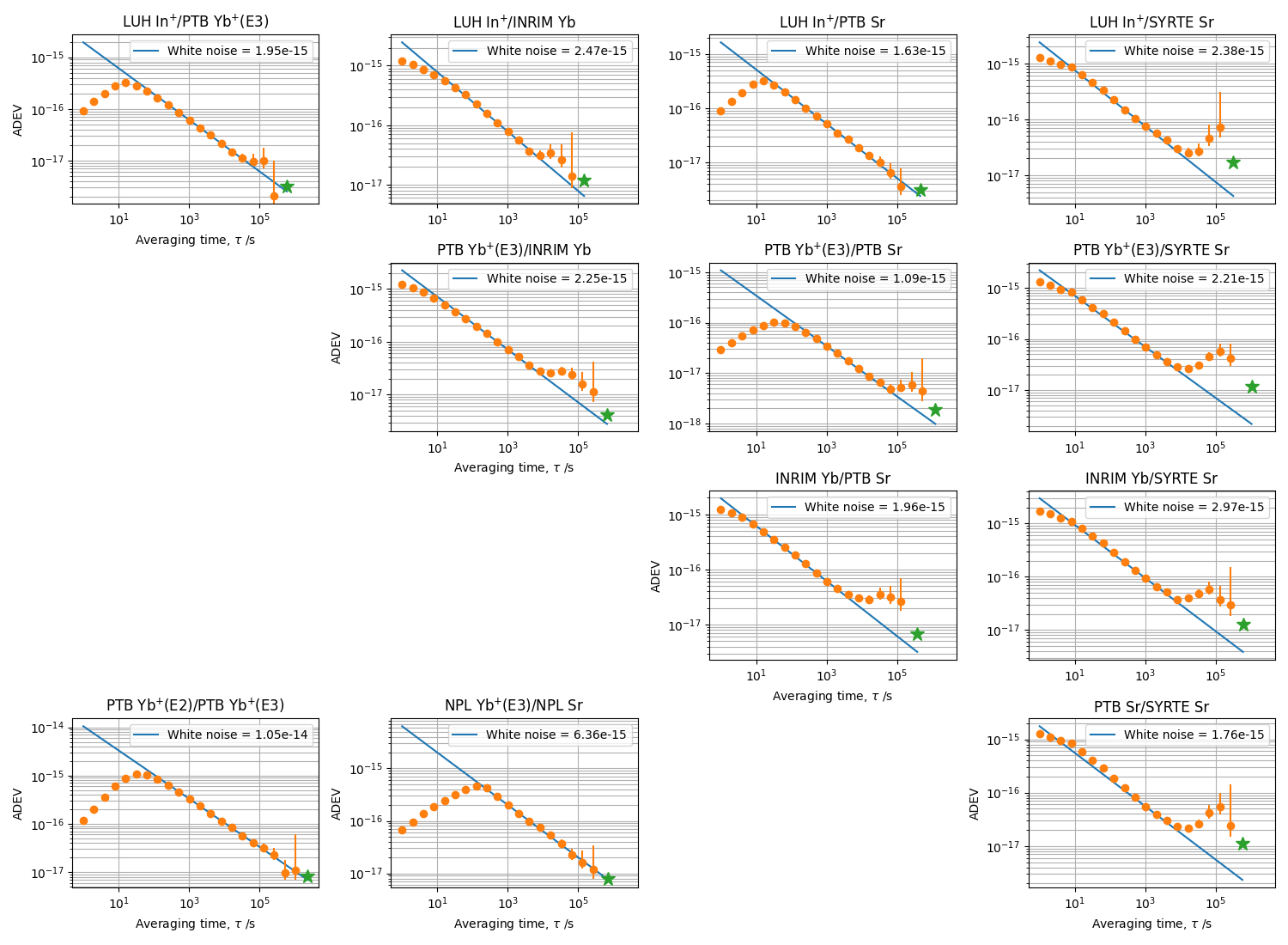}
    \end{subfigure}
    \caption{\textbf{(b)} Overlapping Allan deviations of the clock comparisons. The blue lines show the white frequency instability for each ratio. The value in the legend of each plot reports the value of the white noise Allan deviation at 1 s. The green stars show the statistical uncertainty of each ratio inflated by the Birge ratio at the total measurement time.}
    \label{fig:fibre_link_statistics(b)}
\end{figure}

The data format for clocks allows for the specification of a time-varying systematic uncertainty. In this campaign, the INRIM Yb clock and the PTB Sr clock  reported a time-varying systematic uncertainty owing to changing configurations over the campaign. For the INRIM Yb clock, this ranged from \num{2e-17} to \num{2.8e-17}, while for the PTB Sr clock, it  ranged from \num{2.7e-18} to \num{1.4e-17}. The reported systematic uncertainty for each of these clocks was therefore calculated as the arithmetic mean of the time-varying uncertainties over the uptime of each ratio. Other clocks reported a constant systematic uncertainty for the measurement period.

\section{Calculation of correlation coefficients}
Table \ref{Tab:FibreCorr} reports all the correlation coefficients between the measurements listed in Table 2 of the main text.

\subsection{Correlation from systematic uncertainties}
The correlation coefficient between measurements $i$ and $j$ both involving a clock with systematic uncertainty $u_\text{B}$ was calculated as $r_{ij,B} = \pm u_\text{B}^2 / (u_i u_j)$, where $u_{i}$ and $u_{j}$ are the total uncertainties associated with measurements $i$ and $j$.
The sign  depends on the position of the common clock in the numerator or denominator in the two frequency ratios.
INRIM Yb and PTB Sr reported time-varying systematic uncertainties. In these cases, only the minimum of $u_\text{B}$ was considered correlated between different measurements \cite{S-Margolis2024}.

\subsection{Correlation from statistical uncertainties}
The correlation coefficient between measurements $i$ and $j$ both involving a clock with statistical uncertainty scaling as white noise, $u_A(\tau) = s_0 \tau^{-1/2}$, depends on the averaging time $\tau_i$ and $\tau_j$ of the two measurements and on the overlapping time between the two measurements $\tau_\text{overlap}$ \cite{S-ROCIT_D3}. With $u_{i}$ and $u_{j}$  the total uncertainties associated with measurements $i$ and $j$, the correlation coefficient from the statistical uncertainty is $r_{ij,A} = \pm s_0^2 \tau_\text{overlap}/(\tau_i \tau_j u_i u_j)$. The sign  depends on the position of the common clock in the numerator or denominator in the two frequency ratios.

\subsection{Correlation from extrapolations}
To calculate the level of correlation introduced when a common maser is used to extrapolate across gaps in optical clock data, we generalize the Fourier transform method of calculating extrapolation uncertainties (Ref. \cite{S-Dawkins2007}, Appendix A) to include covariances.
The covariance of the same maser frequency noise $y$ sampled over two weighting functions $w_1$ and $w_2$ (which represent the times when maser data is used for extrapolation) is
\begin{align}
\begin{split}
\sigma^2_{12} & =
\left\langle \int_{-\infty}^{\infty} w_1(t')y(t') dt' \int_{-\infty}^{\infty} w_2(t'')y(t'') dt'' \right\rangle \\
 & = \int_{-\infty}^{\infty} \int_{-\infty}^{\infty} w_1(t')  w_2(t'') \langle y(t')  y(t'')\rangle dt' dt''.
\end{split}
\end{align}
From the definition of the autocorrelation function and the Wiener-Khinchin theorem for the two-sided power spectral density $S^{II}_y(f)$, we have
\begin{equation}
\langle y(t')  y(t'')\rangle = \int_{-\infty}^{\infty} S^{II}_y(f) \exp[i 2 \pi f(t'-t'')]df.
\end{equation}
Substituting this into the previous equation
\begin{align}
\begin{split}
\sigma^2_{12} & =
\int_{-\infty}^{\infty} \int_{-\infty}^{\infty} w_1(t')  w_2(t'') \int_{-\infty}^{\infty} S^{II}_y(f) \exp[i 2 \pi f(t'-t'')]df dt' dt'' \\ & =
\int_{-\infty}^{\infty} \int_{-\infty}^{\infty} \int_{-\infty}^{\infty} w_1(t')  w_2(t'')  S^{II}_y(f) \exp[i 2 \pi f(t'-t'')] dt' dt'' df \\ & =
\int_{-\infty}^{\infty} S^{II}_y(f) \int_{-\infty}^{\infty} w_1(t')   \exp(i 2 \pi f t') dt' \int_{-\infty}^{\infty}  w_2(t'') \exp(- i 2 \pi ft'')  dt'' df \\
& =
\int_{-\infty}^{\infty} S^{II}_y(f) W_1(-f)  W_2(f)  df,
\end{split}
\end{align}
where $W_j(f)$  are the Fourier transforms of $w_j(t)$.
Finally, $W_j(-f) = W_j(f)^*$, where $^*$ denotes the complex conjugate, because $w_j$ are real, so:
\begin{equation}\label{eq:extrapolation_covariance}
\sigma^2_{12} =
\int_{0}^{\infty} S^{I}_y(f) \mathrm{Re}(W_1(f)^*W_2(f)) df,
\end{equation}
where $S^{I}_y(f)$ is the one-sided power spectral density.
Once the covariance is determined, the correlation coefficient can be calculated according to $r_{12} = \sigma_{12} / (u_1 u_2)$, where $u_1$ and $u_2$ are the total uncertainties of the two measurements.

\begin{figure}[bth]
\centering
    \includegraphics[width=0.8\textwidth]{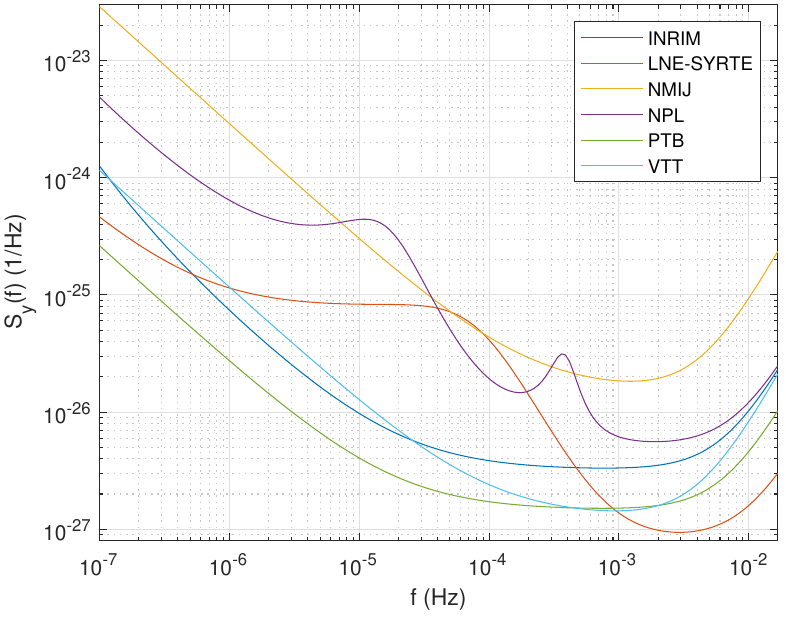}\\
    \caption{Single-sided power spectral density of the maser noise models in Table~\ref{Tab:noisemodels}.}
    \label{fig:maserPSD}
\end{figure}

To evaluate eq. \ref{eq:extrapolation_covariance}, we describe the maser frequency noise $S_y^I(f)$ as a power-law model, with possible dominating bumps due to quasi-periodic behavior modelled by Lorentzian peaks, see Table~\ref{Tab:noisemodels} and Fig.~\ref{fig:maserPSD}.
We note that modelling long-term noise (flicker and random-walk) may introduce long-term correlations even for non-overlapping weighting functions. In our case, neglecting correlations between the two analysis intervals of the GNSS solutions when calculating the ratios affected the extrapolation uncertainties by less than $0.5\%$.

\begin{table}[t]
\caption{Maser noise models. The noise is described as a sum of white phase noise ($h_2$), white frequency noise ($h_0$), flicker frequency noise ($h_{-1}$), and random walk frequency noise ($h_{-2}$), with the respective coefficients in parenthesis,
and, in two cases, Lorentzian peaks that describe quasi-periodic behavior: $S^{I}_y(f) = h_2 f^2 + h_0 + h_{-1}/f + h_{-2}/f^2 + \sum_i A_i/(1 + (f-f_{0,i})^2/\delta f_i^2)$.}
\label{Tab:noisemodels}
\footnotesize\rm
\begin{tabular}{l SSSS SSS SSS}
\toprule
{Institute }&{ $h_2$ }&{ $h_0$ }&{ $h_{-1}$ }&{ $h_{-2}$ }&{ $A_1$ }&{ $f_{0,1}$ }&{ $\delta f_1$ }&{ $A_2$ }&{ $f_{0,2}$ }&{ $\delta f_2$ }\\
&{ $10^{-23}$ }&{ $10^{-27}$ }&{ $10^{-31}$ }&{ $10^{-39}$ }&{ $10^{-26}$ }&{ $10^{-5}$ }&{ $10^{-5}$ }&{ $10^{-26}$ }&{ $10^{-5}$ }&{  $10^{-5}$}\\
 &{ Hz$^{-3}$ }&{ Hz$^{-1}$ }&{ Hz$^{0}$ }&{ Hz }&{ Hz$^{-1}$ }&{ Hz }&{ Hz }&{ Hz$^{-1}$ }&{ Hz }&{ Hz }\\
\midrule
INRIM     & 7.11 & 3.2 & 0.65 & 6.08 & & & & & & \\
LNE-SYRTE & 0.79 & 0.8 & 0.38 & 0.061 & 8 & 2.01 & 8 \\
NMIJ      & 78.9 & 14.8 & 28.9 & 0 & & & & & & \\
NPL       & 7.11 & 5.0 & 4.06 & 6.08 & 40 & 1.16 & 1.2 &  2.5 & 3.7 & 1\\
PTB       & 3.16 & 1.46 & 0.26 & 0.015 & & & & & & \\
VTT       & 7.11 & 1.25 & 1.15 & 0.015 & & & & & & \\
\bottomrule
\end{tabular}
\end{table}

Extrapolation uncertainties also consider the effect of the maser drift rate. In our case, the uncertainties from the drift are small and consequently the resulting correlation coefficients are negligible, ${<}0.001$ in magnitude.

\subsection{IPPP temporal correlation}
To estimate the correlations introduced by different ratios sharing IPPP solutions from common GNSS data, it is necessary to establish the phase noise power spectral density (PSD) of the IPPP link, $S_x$.  This was done by comparing results from IPPP and a White Rabbit fiber link \cite{S-Petit2022}.  The deduced PSD can be modelled by a piecewise polynomial that describes flicker phase noise with a low cut-off frequency $f_\mathrm{l}$, a frequency $f_\mathrm{c}$ above which it falls off as $1/f^2$ (white frequency noise), and a high cut-off frequency $f_\mathrm{h} = 1/(2 \tau_0)$, where $\tau_0 = 30$\;s is the integration time,
\begin{equation} \label{eq:Sx}
S_x(f) = \left\{
    \begin{array}{ll}
    k_{-1} f/f_\mathrm{l}^2 & \mathrm{for}\; f < f_\mathrm{l}, \\
    k_{-1}/f & \mathrm{for}\; f_\mathrm{l} \leq f < f_\mathrm{c}, \\
    k_{-1}f_\mathrm{c}/f^2 & \mathrm{for}\; f_\mathrm{c} \leq f < f_\mathrm{h}, \\
    0 & \mathrm{for}\; f > f_\mathrm{h}.
    \end{array}
    \right.
\end{equation}

\begin{figure}[t]
\centering
    \includegraphics[width=0.9\textwidth]{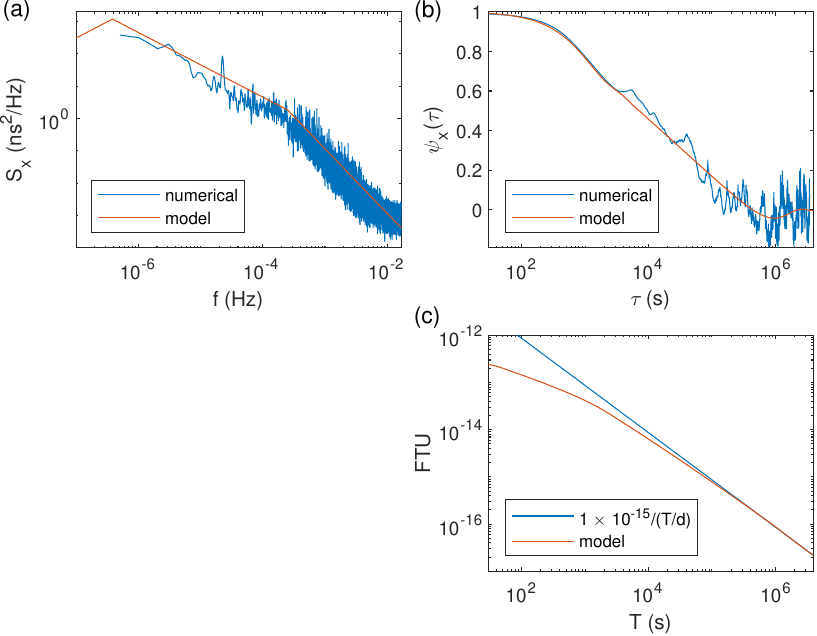}\\
    \caption{(a) Calculated power spectral density $S_x$ (blue) and piecewise model Eq.~\ref{eq:Sx} (red). (b) Normalized autocorrelation functions $\psi_x(\tau) = \Psi_x(\tau)/\Psi_x(0)$: calculated from data (blue) and analytical model Eqs.~(\ref{eq:Psi}--\ref{eq:Psi0}) (red). (c) Frequency transfer uncertainty from model (red) compared to the $1\times 10^{-15}/(T/\mathrm{d})$ limit (blue).}
    \label{fig:IPPPcorr}
\end{figure}

The phase autocorrelation function $\Psi_x(\tau) = \langle x(t) x(t+\tau) \rangle$, where $\tau$ is the lag, was then calculated analytically as the Fourier transform of the piecewise $S_x$ model, yielding
\begin{align}
\Psi_x(\tau) = k_{-1} & \bigg\{ \frac{\cos{(2\pi f_\mathrm{l}\tau)} - 1 +
    2\pi f_\mathrm{l}\tau \sin{(2\pi f_\mathrm{l}\tau})}{(2\pi f_\mathrm{l}\tau)^2}
    + \Ci{(2\pi f_\mathrm{c}\tau)} - \Ci{(2\pi f_\mathrm{l}\tau)} \nonumber \\
    & \; \; + \cos{(2\pi f_\mathrm{c}\tau)} - \frac{f_\mathrm{c}}{f_\mathrm{h}} \cos{(2\pi f_\mathrm{h}\tau)} + 2\pi f_\mathrm{c}\tau
    \big[\Si{(2\pi f_\mathrm{c}\tau)} - \Si{(2\pi f_\mathrm{h}\tau)}\big] \bigg\}, \label{eq:Psi} \\
\Psi_x(0) = k_{-1} & \left[ \frac{3}{2} + \ln{\frac{f_\mathrm{c}}{f_\mathrm{l}}
- \frac{f_\mathrm{c}}{f_\mathrm{h}}} \right], \label{eq:Psi0}
\end{align}
where the cosine and sine integral functions are defined as
\begin{equation}
\Ci{(x)} = -\int_x^\infty \frac{\cos{y}}{y} \mathrm{d}y \quad \text{and} \quad \Si{(x)} = \int_0^x \frac{\sin{y}}{y} \mathrm{d}y, \quad x>0,
\end{equation}
respectively. For the relevant parameter values, no satisfying approximation was found for the  autocorrelation function, so the full expression was used. The low cut-off frequency in \eqref{eq:Sx} was chosen as $f_\mathrm{l} = 1/(\SI{30}{d})$ to give the best agreement between \eqref{eq:Psi} and the autocorrelation function numerically calculated from the IPPP-fiber link data, see Fig.~\ref{fig:IPPPcorr}(b).

The IPPP frequency transfer uncertainty is $\text{FTU} = \sqrt{\Var{y}}$, where we use the expression for the mean frequency, $y = (x_{\mathrm{end}} - x_{\mathrm{start}})/\tau$, to write the variance as $\Var{y} = 2/\tau^2 \left[ \Psi_x(0) - \Psi_x(\tau) \right]$. The amplitude of the $S_x$ model in Fig.~\ref{fig:IPPPcorr}(a) was slightly increased compared to a fit to the numerical $S_x$ to make the FTU agree with the conservative estimate of $1\times 10^{-15}/(T/\mathrm{d})$ for intervals above 1 day, see Fig.~\ref{fig:IPPPcorr}(c). This corresponds to $k_{-1} = 4.6\times 10^{-22}\;\mathrm{s^2/Hz}$.

The PSD $S_x$ also has a peak at a frequency corresponding to the GPS satellite orbit time (11\;h 58\;min), which gives rise to oscillations in the autocorrelation function. As these features vary between receivers, they were neglected from the model and therefore we truncate the autocorrelation function at its first zero around \SI{5.06}{d}.

The covariance $\langle y_i y_j \rangle$ between the mean frequencies of two IPPP intervals can be calculated  using $\Psi_x(\tau)$ similarly to how the variance was evaluated above.
This analysis showed that by neglecting the correlation between the two analysis intervals when evaluating the ratios, we slightly overestimated the total link uncertainty for intervals separated by less than 5 d; in the worst case by $4.5\%$ for intervals separated by only 1.1\;d.

\newpage
\begin{small}
\begin{longtable}{l@{\ $=$\ }S[table-format=+2.6] l@{\ $=$\ }S[table-format=+2.6] l@{\ $=$\ }S[table-format=+2.6] l@{\ $=$\ }S[table-format=+2.6]}

\caption{Correlation coefficients for the measurements in Table~2 in the main text. The coefficients $r(1,2)$ through $r(11,12)$ are between fiber and local ratios, while $r(13,14)$ onwards are between GNSS ratios.} \label{Tab:FibreCorr}\\

\toprule
\endfirsthead

\multicolumn{8}{c}%
{{\tablename\ \thetable{} (continued)}} \\
\toprule
\endhead

\bottomrule
\endfoot

\bottomrule
\endlastfoot

$r(\text{1,2})$ & 0.132	& $r(\text{1,3})$ & 0.507	& $r(\text{1,4})$ & 0.211	& $r(\text{1,5})$ & 0.060\\
$r(\text{1,6})$ & -0.080	& $r(\text{1,8})$ & -0.347	& $r(\text{1,9})$ & -0.111	& $r(\text{2,3})$ & 0.145\\
$r(\text{2,4})$ & 0.073	& $r(\text{2,6})$ & 0.854	& $r(\text{2,9})$ & 0.011	& $r(\text{2,10})$ & -0.865\\
$r(\text{2,11})$ & -0.624	& $r(\text{2,12})$ & 0.011	& $r(\text{3,4})$ & 0.200	& $r(\text{3,8})$ & 0.385\\
$r(\text{3,10})$ & 0.094	& $r(\text{3,12})$ & -0.097	& $r(\text{4,6})$ & 0.011	& $r(\text{4,9})$ & 0.725\\
$r(\text{4,10})$ & -0.011	& $r(\text{4,11})$ & 0.443	& $r(\text{4,12})$ & 0.730	& $r(\text{5,6})$ & -0.014\\
$r(\text{5,8})$ & -0.055	& $r(\text{5,9})$ & -0.016	& $r(\text{6,8})$ & 0.077	& $r(\text{6,9})$ & 0.040\\
$r(\text{6,10})$ & -0.940	& $r(\text{6,11})$ & -0.699	& $r(\text{6,12})$ & 0.012	& $r(\text{8,9})$ & 0.108\\
$r(\text{8,10})$ & 0.132	& $r(\text{8,12})$ & -0.159	& $r(\text{9,10})$ & -0.012	& $r(\text{9,11})$ & 0.543\\
$r(\text{9,12})$ & 0.848	& $r(\text{10,11})$ & 0.691	& $r(\text{10,12})$ & -0.046	& $r(\text{11,12})$ & 0.530\\
\midrule
$r(\text{13,14})$ & 0.874	& $r(\text{13,15})$ & 0.740	& $r(\text{13,16})$ & -0.015	& $r(\text{13,17})$ & -0.007\\
$r(\text{13,18})$ & 0.758	& $r(\text{13,19})$ & -0.030	& $r(\text{13,20})$ & 0.671	& $r(\text{13,21})$ & 0.815\\
$r(\text{13,22})$ & -0.147	& $r(\text{13,23})$ & -0.007	& $r(\text{13,27})$ & -0.125	& $r(\text{13,28})$ & 0.006\\
$r(\text{13,30})$ & -0.114	& $r(\text{13,31})$ & 0.006	& $r(\text{13,33})$ & -0.060	& $r(\text{13,34})$ & 0.004\\
$r(\text{13,36})$ & 0.201	& $r(\text{13,37})$ & 0.136	& $r(\text{13,38})$ & -0.004	& $r(\text{14,15})$ & 0.685\\
$r(\text{14,16})$ & 0.175	& $r(\text{14,18})$ & 0.720	& $r(\text{14,20})$ & 0.564	& $r(\text{14,21})$ & 0.758\\
$r(\text{14,22})$ & -0.125	& $r(\text{14,24})$ & -0.088	& $r(\text{14,25})$ & -0.198	& $r(\text{14,27})$ & -0.239\\
$r(\text{14,28})$ & -0.086	& $r(\text{14,29})$ & -0.115	& $r(\text{14,30})$ & -0.106	& $r(\text{14,33})$ & -0.057\\
$r(\text{14,36})$ & 0.182	& $r(\text{14,37})$ & 0.127	& $r(\text{15,17})$ & 0.421	& $r(\text{15,18})$ & 0.538\\
$r(\text{15,20})$ & 0.508	& $r(\text{15,21})$ & 0.643	& $r(\text{15,22})$ & -0.116	& $r(\text{15,24})$ & 0.391\\
$r(\text{15,26})$ & -0.408	& $r(\text{15,27})$ & -0.108	& $r(\text{15,30})$ & -0.440	& $r(\text{15,31})$ & -0.356\\
$r(\text{15,32})$ & -0.360	& $r(\text{15,33})$ & -0.047	& $r(\text{15,36})$ & 0.145	& $r(\text{15,37})$ & 0.138\\
$r(\text{16,17})$ & 0.050	& $r(\text{16,19})$ & 0.040	& $r(\text{16,20})$ & -0.009	& $r(\text{16,22})$ & 0.039\\
$r(\text{16,23})$ & 0.065	& $r(\text{16,24})$ & -0.369	& $r(\text{16,25})$ & -0.354	& $r(\text{16,27})$ & -0.289\\
$r(\text{16,28})$ & -0.638	& $r(\text{16,29})$ & -0.435	& $r(\text{16,31})$ & -0.038	& $r(\text{16,34})$ & -0.018\\
$r(\text{16,36})$ & -0.041	& $r(\text{16,38})$ & 0.072	& $r(\text{17,19})$ & 0.011	& $r(\text{17,20})$ & -0.004\\
$r(\text{17,22})$ & 0.008	& $r(\text{17,23})$ & 0.032	& $r(\text{17,24})$ & 0.843	& $r(\text{17,26})$ & -0.600\\
$r(\text{17,28})$ & -0.029	& $r(\text{17,30})$ & -0.579	& $r(\text{17,31})$ & -0.878	& $r(\text{17,32})$ & -0.755\\
$r(\text{17,34})$ & 0.000	& $r(\text{17,36})$ & -0.001	& $r(\text{17,38})$ & 0.023	& $r(\text{18,19})$ & 0.173\\
$r(\text{18,20})$ & 0.256	& $r(\text{18,21})$ & 0.654	& $r(\text{18,22})$ & -0.063	& $r(\text{18,25})$ & 0.140\\
$r(\text{18,26})$ & 0.094	& $r(\text{18,27})$ & -0.056	& $r(\text{18,30})$ & -0.041	& $r(\text{18,33})$ & -0.149\\
$r(\text{18,34})$ & -0.046	& $r(\text{18,35})$ & -0.112	& $r(\text{18,36})$ & 0.113	& $r(\text{18,37})$ & 0.067\\
$r(\text{19,20})$ & -0.010	& $r(\text{19,22})$ & 0.031	& $r(\text{19,23})$ & 0.025	& $r(\text{19,25})$ & 0.403\\
$r(\text{19,26})$ & 0.220	& $r(\text{19,28})$ & -0.017	& $r(\text{19,31})$ & -0.005	& $r(\text{19,33})$ & -0.256\\
$r(\text{19,34})$ & -0.362	& $r(\text{19,35})$ & -0.321	& $r(\text{19,36})$ & -0.016	& $r(\text{19,38})$ & 0.015\\
$r(\text{20,21})$ & 0.541	& $r(\text{20,22})$ & -0.218	& $r(\text{20,23})$ & -0.006	& $r(\text{20,27})$ & -0.195\\
$r(\text{20,28})$ & 0.020	& $r(\text{20,30})$ & -0.177	& $r(\text{20,31})$ & 0.011	& $r(\text{20,33})$ & -0.131\\
$r(\text{20,34})$ & 0.083	& $r(\text{20,36})$ & 0.273	& $r(\text{20,37})$ & 0.156	& $r(\text{20,38})$ & -0.029\\
$r(\text{21,22})$ & -0.117	& $r(\text{21,23})$ & 0.256	& $r(\text{21,27})$ & -0.109	& $r(\text{21,29})$ & 0.292\\
$r(\text{21,30})$ & -0.099	& $r(\text{21,32})$ & 0.194	& $r(\text{21,33})$ & -0.052	& $r(\text{21,35})$ & 0.346\\
$r(\text{21,36})$ & 0.169	& $r(\text{21,37})$ & 0.325	& $r(\text{21,38})$ & 0.198	& $r(\text{22,23})$ & 0.027\\
$r(\text{22,27})$ & 0.816	& $r(\text{22,28})$ & -0.030	& $r(\text{22,30})$ & 0.531	& $r(\text{22,31})$ & -0.006\\
$r(\text{22,33})$ & 0.657	& $r(\text{22,34})$ & -0.016	& $r(\text{22,36})$ & -0.714	& $r(\text{22,37})$ & -0.622\\
$r(\text{22,38})$ & 0.024	& $r(\text{23,28})$ & -0.084	& $r(\text{23,29})$ & 0.573	& $r(\text{23,31})$ & -0.042\\
$r(\text{23,32})$ & 0.426	& $r(\text{23,34})$ & -0.021	& $r(\text{23,35})$ & 0.308	& $r(\text{23,36})$ & -0.033\\
$r(\text{23,37})$ & 0.374	& $r(\text{23,38})$ & 0.780	& $r(\text{24,25})$ & 0.123	& $r(\text{24,26})$ & -0.572\\
$r(\text{24,27})$ & 0.108	& $r(\text{24,28})$ & 0.280	& $r(\text{24,29})$ & 0.207	& $r(\text{24,30})$ & -0.536\\
$r(\text{24,31})$ & -0.743	& $r(\text{24,32})$ & -0.680	& $r(\text{25,26})$ & 0.165	& $r(\text{25,27})$ & 0.259\\
$r(\text{25,28})$ & 0.103	& $r(\text{25,29})$ & 0.219	& $r(\text{25,33})$ & -0.191	& $r(\text{25,34})$ & -0.181\\
$r(\text{25,35})$ & -0.301	& $r(\text{26,30})$ & 0.520	& $r(\text{26,31})$ & 0.456	& $r(\text{26,32})$ & 0.520\\
$r(\text{26,33})$ & -0.138	& $r(\text{26,34})$ & -0.028	& $r(\text{26,35})$ & -0.132	& $r(\text{27,28})$ & 0.160\\
$r(\text{27,29})$ & 0.192	& $r(\text{27,30})$ & 0.498	& $r(\text{27,33})$ & 0.601	& $r(\text{27,36})$ & -0.640\\
$r(\text{27,37})$ & -0.537	& $r(\text{28,29})$ & 0.402	& $r(\text{28,31})$ & 0.097	& $r(\text{28,34})$ & 0.028\\
$r(\text{28,36})$ & 0.082	& $r(\text{28,38})$ & -0.165	& $r(\text{29,32})$ & 0.366	& $r(\text{29,35})$ & 0.407\\
$r(\text{29,37})$ & 0.333	& $r(\text{29,38})$ & 0.524	& $r(\text{30,31})$ & 0.508	& $r(\text{30,32})$ & 0.496\\
$r(\text{30,33})$ & 0.387	& $r(\text{30,36})$ & -0.354	& $r(\text{30,37})$ & -0.340	& $r(\text{31,32})$ & 0.748\\
$r(\text{31,34})$ & 0.003	& $r(\text{31,36})$ & 0.019	& $r(\text{31,38})$ & -0.070	& $r(\text{32,35})$ & 0.222\\
$r(\text{32,37})$ & 0.214	& $r(\text{32,38})$ & 0.372	& $r(\text{33,34})$ & 0.098	& $r(\text{33,35})$ & 0.152\\
$r(\text{33,36})$ & -0.266	& $r(\text{33,37})$ & -0.273	& $r(\text{34,35})$ & 0.144	& $r(\text{34,36})$ & 0.071\\
$r(\text{34,38})$ & -0.093	& $r(\text{35,37})$ & 0.278	& $r(\text{35,38})$ & 0.175	& $r(\text{36,37})$ & 0.564\\
$r(\text{36,38})$ & -0.084	& $r(\text{37,38})$ & 0.311	 \\

\end{longtable}
\end{small}


\end{document}